# Mid-Infrared Microscopy for Label-Free Digital Staining: Initial Clinical Assessment

L. Duraffourg[1*], H. Borges[1], M. Fernandes[1], M. Beurrier-Bousquet[1], J. Baraillon[1], B. Taurel[1], J. Le Galudec[1], K. Vianey[2], C. Maisin[2], L. Samaison[1], F. Staroz[1], M. Dupoy[1]

[1]ADMIR, 137 rue Mayoussard, Bat C, 38430 Moirans, France

[2] Cypath-RB, 201 Route de Genas, 69100 Villeurbanne, France

*Corresponding Author: laurent.duraffourg@admir-analysis.com

Keywords: Digital pathology, Infrared Spectroscopy, digital staining, Microscopy, Deep Learning

Abstract

Infrared (IR) microscopy shows substantial potential for label-free tissue imaging in anatomic pathology, providing rich biochemical contrast. However, existing IR imaging technologies are constrained by slow acquisition speeds and limited spatial resolution. Here, we present a rapid, large-field bimodal imaging platform that integrates conventional brightfield microscopy with a lensless IR imaging scanner, enabling whole-slide IR image stack acquisition in minutes. Using a dedicated deep learning model, we implement an optical H&E staining strategy based on subcellular morpho-spectral fingerprinting. This approach achieves high-resolution visualization of tissue architecture with an effective spatial resolution of 500 nm, without chemical staining.

Quantitative metrics, including PSNR (~24), MS-SSIM (~0.82), and LPIPS (~0.22), validate the model's ability to accurately reproduce both the contrast and morphology of cellular structures. Additionally, an initial clinical evaluation on 110 regions of interest within prostate tissue sections demonstrates equivalence between our digital IR-based AI staining and conventional chemical staining, both in image quality and Gleason grading.

## Introduction

Anatomic pathology is the branch of medical science that studies the causes, mechanisms, and effects of diseases. It focuses on observing tissues, cells, and bodily fluids to understand abnormalities, diagnose diseases, and predict their progression. Pathology serves as a cornerstone of modern medicine, providing insights that guide diagnosis, therapy, and prognosis.

The diagnosis process flow involves collecting and processing tissue samples through fixation, dehydration, embedding, and thin sectioning, before staining with Hematoxylin and Eosin (H&E). Pathologists then examine the slides under a microscope to identify cellular and structural abnormalities.

The H&E staining is in the first step in 95% of cancer diagnosis (Cotran et al., 1999). Complementary tests can be done if necessary. In this case, pathologists take back other sections and redo sample preparations for special staining, which highlights specific components (e.g., Periodic Acid-Schiff

(PAS) for carbohydrates, Masson's Trichrome for collagen…), or for immunohistochemistry (IHC). If necessary, molecular pathology testing are finally performed (Polymerase Chain Reaction – PCR, DNA or RNA sequencing).

Today, the pathologist community is facing three major challenges, namely the ever-increasing demand for tests, the increasing amount of information per test, and the fact that diagnostic error has to be minimized. Moreover, laboratories face a critical shortage of both pathologists and technical personnel. This challenge comes at a pivotal moment, as medicine shifts toward precision and personalized care, an approach that demands a nuanced understanding of cancer's intricate mechanisms, from the genetic and proteomic levels to the metabolomic landscape (Hackshaw et al., 2020; Senevirathna et al., 2021). For instance, the heterogeneity of cancers has been reported at the expression level of numerous protein markers, at the genetic, epigenetic and transcriptional levels, and more recently at the metabolic and tumor microenvironment levels. Many of these biological characteristics are spatially heterogeneously distributed. Taking account of the different levels of biological heterogeneity (proteinic, molecular, tissue and cellular, metabolic, spatial & temporal) (Gerdes et al., 2014; Kashyap et al., 2022) is essential for improving patient stratification and personalizing treatments.

In this context, the emergence of digital pathology is a natural response to the need for greater efficiency (Hanna and Ardon, 2023) and cancer mechanism understanding. It enables slide image retrieval, report integration, as well as large-scale data mining and research through structured, digitized datasets. Moreover, digital pathology enables the application of artificial intelligence (AI) and quantitative image analysis tools, which may improve diagnostic accuracy, tumor grading, and biomarker quantification. AI-based prognostic models, for example, would offer objective and reproducible assessments that enhance clinical decision-making. Beyond the analysis of known biomarkers, digital pathology holds the potential to accelerate the discovery, validation, and clinical adoption of novel biomarkers, thereby advancing the field of precision medicine.

However, the widespread implementation of digital pathology faces several important challenges and limitations (Sirintrapun, 2025). Some are operational, including the need for high-capacity data storage, substantial financial investment, and the establishment of universal standards and interoperability frameworks. Others are technical, particularly concerning the generalizability and validation of AI algorithms across real-world, heterogeneous datasets. These include variations in scanner hardware, staining protocols across laboratories and the diversity of tissue types. Digitization of complex or special stains also remains technically demanding.

Label-free imaging has recently emerged as a promising way to overcome these technical challenges, thereby accelerating and streamlining the pathology workflow and the integration of IA. In particular, virtual staining of tissue sections that simulates traditional histological staining methods is a popular approach (Bai et al., 2023; Yilmaz et al., 2023; Szulczewski et al., 2024). Once the image is acquired using a slide scanner or a digital microscope, image processing algorithms or artificial intelligence models (Latonen et al., 2024), apply a virtual "stain" that simulates the visual appearance of a conventional chemical staining. Still, virtual staining is facing several significant problems. One relates to the intrinsic variability of AI staining models. Generating accurate nuclei staining remains particularly challenging, as models often either underproduce or overproduce nuclei as it can be noticed in (Koivukoski et al., 2023). Cellular or extra cellular details quality, for instance membrane detail, remains also suboptimal. In addition, the algorithms are constrained to morphological feature analysis and lack the capacity to identify cell types beyond their training data, especially when confronted with abnormal or unseen phenotypes.

To tackle the problems, it has been suggested to replace the traditional brightfield imaging, with alternative microscopy techniques. Approaches such as quantitative Phase Imaging (QPI)[1] or ptychography have been suggested as ways to produce more contrasted images (Rivenson et al., 2019; Wang et al., 2024). However, changing the light source itself could unlock access to a wide range of information.

Indeed, light-tissue interaction, such as absorption and emission, depends intrinsically on the tissue type and cellular elements. These spectral features are intimately related to the macromolecules forming the tissue and can be used to help the AI-models to distinguish different cellular structure. This simple idea has been suggested by several research groups and startups. Among the most promising way, autofluorescence imaging (Epi-illumination) seems to be one of the most advanced (Bai et al., 2023). In this case multispectral image stack corresponding to intrinsic fluorescence of collagen, NADH, flavins and a few others targets are acquired when tissue is irradiated with UV or blue-light. This can be done with standard fluorescent slide scanners with dedicated filters at optical magnification from 20x to 40x, enabling a spatial resolution similar to current BF-scanner resolution. This approach is now the technological baseline of the startup Pictor Labs, which is developing GAN-based AI models for digital staining through style transfer model (Bai et al., 2022). A similar method based on UV fluorescence has also been suggested. The imaging modality uses a simple grayscale reflectance imaging system at low magnification that primarily highlights nuclei (at 365nm-excitation wavelength). This approach is suggested by startups like MUSE microscopy or Vivascope. The main applications are in dermatology or at the surgery block for analyzing thick or opaque samples. Such techniques offer lateral spatial resolution that ranges from 250nm to 1µm, but suffer from slow acquisition rate. Imaging of 1 cm² can take up to several tens of minutes, making these techniques incompatible with a clinical workflow.

Similarly, Raman and Infrared spectroscopies have been explored as approaches for digital staining. Despite being used by the startup Lightcore Technologies for a single test in digital staining (Sarri et al., 2019), Raman spectroscopy remains quite confidential for this specific application. On the other hand, IR spectroscopy has extensively been studied as a powerful tool to both perform a digital H&E staining and efficient tissue segmentation (Beć et al., 2020). It has been used since the 1980s in chemistry and physics to analyze materials across the 4000–400 cm$^{-1}$ range. It is an analytical technique that exploits the interaction of infrared radiation with matter to study and identify chemical compounds. Molecules absorb specific IR frequencies, causing vibrational transitions in their molecular bonds. These absorption patterns are characteristic of a molecule's structure, making it possible to identify functional groups and determine molecular composition. IR spectroscopy can characterize organic and inorganic compounds in various states (solid, liquid, gas) and is applied in many fields (materials science, pollutant detection, pharmaceutical analysis…)

More recently, IR microscopy has emerged as a versatile analytical tool that combines IR spectroscopy with optical microscopy to enable spatially resolved chemical analysis. It provides detailed insights into the composition and distribution of materials at microscopic scales. Infrared imaging microscopy has then naturally emerged as a valuable tool for the investigation of tissue sections and cell cultures. It provides a direct non-destructive method for chemical identification without the need of additional chemicals (no reagent, no labelling, no staining). The infrared beam passing through a sample reveals the chemical bonds inside proteins, lipids, nucleic acids and more globally primary metabolites (Finlayson et al., 2019; Lorenz-Fonfria, 2020).

---

[1] In QPI techniques we include Phase contrast or differential interference contrast (DIC).

Fourier Transform InfraRed (FTIR) microscopes have already demonstrated their ability to image tissues and cells with diagnosis performances that are comparable to conventional approaches such as H&E staining or IHC (Hughes and Baker, 2016). The approach has therefore been validated, including for identifying the grade of certain cancers (Amrania et al., 2018; Ellis et al., 2023).

Such an approach has nevertheless long been limited by the low-power of available polychromatic IR sources. This has made IR microscopes extremely slow for tissue imaging, with acquisition times of several hours for areas of a few mm² (Balan et al., 2019; Finlayson et al., 2019). Moreover, spatial resolution remains at best close to cell size (~5-10µm). Recent improvements in uncooled infrared imagers now enable camera-speed acquisitions, without the need for bulky and cryogenic cooling systems (Niklaus et al., 2007). These advancements open new possibilities for spectral imaging in the IR range. With the advent of high-brilliance IR light sources, such as Quantum Cascade Lasers (QCLs) (Faist et al., 1994; Kröger et al., 2014; Vitiello et al., 2015), several optical setups have been suggested and even commercialized this last decade, namely: optical infrared-transmission microscopy, optical photothermal infrared microscopy, and fluorescence-detected Mid-infrared (Mid-IR) photothermal microscopy (Shi et al., 2020). Such approaches have reached spatial resolution around 2µm (Zhang et al., 2016; Schnell et al., 2020; Xia et al., 2022). Despite compelling demonstrations, these approaches are limited by both low spatial resolution, falling short of pathologists' requirements for nuclear-level detail and slow acquisition rates (exceeding 10 minutes per Whole Slide Image, WSI). These two technological limitations currently confine IR microscopy to the proof-of-concept phase, where its use is largely restricted to correlating IR absorption maps with immunohistochemical biomarker analyses. Without a paradigm shift, the path to routine use of IR spectroscopy for cancer segmentation in diagnostic and prognostic workflows may remain long and complex.

This study presents a digital H&E staining method based on a bimodal microscopy approach, coupled with an AI model that leverages both morphological features and IR spectral signatures, the 'morpho-spectral fingerprint', to enhance diagnostic precision in anatomic pathology. This technique may represent the paradigm shift needed, and our results demonstrate its potential to standardize IR spectroscopy for clinical use. We introduce a novel multimodal imaging approach that combines an infrared lens-free optical scanner (Mathieu et al., 2021), enabling wide field-of-view imaging at high speed (1min/1cm²), with conventional visible-light microscopy techniques to reach a 500nm-spatial resolution.

This paper presents a comprehensive analysis of our approach across three aspects: instrumentation specifications, mathematical evaluation of our deep learning model (metrics), and biostatistical assessment of preclinical prostate tissue data. The results are compared with autofluorescence-based method, which represents the current state of the art.

# 1 Materials and methods

## 1.1 Biological samples

For this demonstration, we focused our work on prostate cancer. The study included fifteen patients, from which fifteen 3µm-thick tissue sections were prepared. The unstained formalin-fixed paraffin-embedded (FFPE) tissue sections were deposited on microscopy slides made from Calcium Fluoride ($CaF_2$). These slides are transparent in the visible and mid-IR ranges. Detailed information about the 15 biological samples are enclosed in supplementary information (SI).

Once prepared, unstained slides are imaged in mid-IR light, and in visible brightfield. Once acquisitions are completed, all tissues sections undergoes H&E histochemical staining. These stained slides are once again acquired in BF. Thus, for each tissue section, three images are acquired: unstained infrared, unstained brightfield, and H&E stained brightfield. This last image serve as ground-truth for our deep learning (DL) model.

### 1.2 Image acquisition protocol and data set construction

The overall workflow for image acquisition and processing is illustrated in the Figure 1. The acquisition procedure was as follows: (i) Infrared WSI (IR-WSI) were obtained from unstained tissue sections; (ii) BF images were acquired from multiple regions of interest (BF-ROIs) within the same sections; and (iii) following H&E staining of these sections, the corresponding ROIs were re-imaged under brightfield microscopy (BF-H&E-ROIs). The BF-H&E-ROIs are first aligned to the BF-unlabeled ROIs through a custom automated registration pipeline, followed by registration of the IR-ROIs to the same BF-ROIs, achieving near-pixel-level alignment (Figure 1). The automated pipeline incorporates a rough eosin style-transfer model, used to convert BF-ROIs into images similar to BF-H&E-ROIs. While the results are of sufficient quality for pathological applications, they still allow image registration through classical computer-vision algorithms (see section 1.2.3).

The IR-ROIs are automatically retrieved from homemade method and algorithms that localizes the BF-ROI positions onto the IR-WSI and extracts them. Prior to the IR-ROI extraction, the 11µm-spatial resolution of IR images is enhanced to 8 µm via an interpolation algorithm. Finally, areas with too strong deformations or optical artefacts are removed through exclusion masks. ROIs are then split into patches for deep learning training. The acquisition protocol was realized on the 15 prostate tissue sections. At the end, we got 118 602 triplets of patch images:

- 16x16 pixel IR-patch images – 16 channels,
- 256x256 BF-patch images – 3 channels
- and 256x256 BF-H&E-patch images – 3 channels

In the following sections, we detailed the optical platforms and the image post-processing for generating the image data sets.

### 1.2.1 Infrared optical setup

We have developed a custom-built, multi-wavelength mid-IR illumination system, integrated with a wide-field, lensless imaging setup. The IR platform has been built upon two main parts, Figure 2(A):

- The IR source, Figure 2 (B) – It incorporates sixteen discrete QCLs (Faist et al., 1994; Bird and Baker, 2015; Spott et al., 2016; Coutard et al., 2020), each set to emit at wavelengths corresponding to the rovibrational transitions of chemical bonds present in essential biomolecular classes, including nucleic acids (DNA, RNA), proteins, and carbohydrates (Zhizhina and Oieinik, 1972; Barth, 2007; Baker et al., 2014). The laser beams are combined using automated motorized brushless rotation and translation stages, and are expanded using a ×3 afocal optical system – consisting of two gold off-axis parabolic mirrors with focal lengths of 25 mm and 75 mm – to achieve uniform illumination over a 3mm² area, see Figure 2 (C). The illumination uniformity is characterized by a contrast $C = \sigma_I/\Gamma$ of less than 5%, where $\sigma_I$ and $\Gamma$ represent the standard deviation and mean intensity across the illuminated field, respectively. Each QCL is individually controlled by custom-designed electronic boards that provide thermal regulation via a proportional-integral-derivative (PID) feedback loop and sequential activation to image the sample at chosen wavelengths, delivering an average optical power between 20 and 30mW.

- The IR image acquisition setup – The IR transmission images are realized using an uncooled bolometer array (80 × 80 pixels, 34µm pitch) with a dedicated packaging (Yon et al., 2014) that allows the protective cover to be removed to make it broadband[2] and a motorized microscopy XY-stage (MLS203 Thorlabs), Figure 2 (D), providing an image at each wavelength, in grey levels. The infrared camera is placed close to the sample at <0.5mm distance typically. The frame rate is 30Hz. WSIs are reconstructed from individual image tiles, each corresponding to the camera's field of view (FOV), with a 20% overlap between adjacent tiles, in both directions. A super resolution by sub-pixel shifting approach is used to increase the IR spatial resolution (11µm effective pixel size) (Cheol Park et al., 2003). The spatial resolution is directly set by the camera pitch and the acquisition method as the system does not include any objective. At the end of the acquisition, we get "tiled" WSI as shown in Figure 3 (A). The total acquisition time across all wavelengths is on the order of a few minutes per square centimeter.

Following acquisition, a dedicated post-processing pipeline, see Figure 3 (A), was applied to the raw infrared images at each wavelength ($I_{raw\_\lambda i}$) to generate bias-corrected transmission images ($T_{\lambda i}$). The equation (1) summarizes the correction:

$$T_{\lambda i} = \frac{I_{raw\_\lambda i} - I_{offset}}{I_{bkg\_\lambda i} - I_{offset}} \quad (1)$$

The offset image $I_{offset}$ is obtained by averaging 5 to 15 successive frames, in absence of illumination (see the correction effect in Figure 3 (B)), while the background image $I_{bkg\_\lambda i}$ at the wavelength $\lambda_i$ is computed by averaging several IR image regions that do not contain tissue.

To accurately isolate background areas, we developed an automated segmentation algorithm that excludes both the tissue sample and slide artifacts (dust, defaults...), see the correction effect in Figure 3 (C). The resulting transmission image $T_{LR}$ is shown in Figure 3 (D). To smooth the tiling effect, a gradient leveling algorithm (Tweel et al., 2024), that uses overlap areas between adjacent tiles, was applied on $T_{LR}$ to achieve a proper low resolution transmission image $T_{GL}$, see Figure 3 (E). This procedure was applied on each sub-pixel shifted images. These were finally averaged on a higher resolved grid to build the high resolution transmission WSI $T_{HR}$ (11µm-equivalent pixel). For each IR-WSI, an histogram is superimposed (see Figure 3 (A–F)). The histograms, which represent the pixel distribution according to the grey level, display clearly two pixel classes corresponding to the background and the tissue. The signal to background ratio is notably enhanced during the whole process. In addition the spatial details are clearly better contrasted after oversampling, as showed in the different zoomed areas below the WSIs.

Finally, we got a stack of 16 oversampled IR-WSIs, corresponding to the 16 IR-wavelength used to construct the IR images (see Figure 1) for the training and test of our digital staining model. The spatial resolution after the oversampling was estimated at 11 µm. To simplify deep learning training, images were then upscaled by interpolation to reach a 8µm-pixel size (see Figure 1).

---

[2] Broadband means that the camera maintains consistent sensitivity across the entire useful infrared wavelength range.

### 1.2.2 Bright field optical setup

The BF image acquisitions were performed with a commercial Zeiss microscope (Axioscope 5) with a 20× $NA_{obj}$ 0.45 objective and a CMOS camera (3.45µm pitch, 2464 × 2056 pixels) with an effective pixel size of 274nm in the object-space. The condenser numerical aperture $NA_{cond}$ was adjusted to match that of the objective. The spatial resolution set by the optical diffractive limit in equation (2) is 705nm for a wavelength λ=520nm, which in principle allows the nucleus of a cell to be resolved when cell elements are enough contrasted.

$$\theta = \frac{1.22 \times \lambda}{NA_{obj} + NA_{cond}} \quad (2)$$

The microscope has been equipped with a two-axis motorized microscopy stage (MLS203, Thorlabs) to enable automated scanning of tissue sections within the camera's FOV, following an S-pattern trajectory with 20% overlap between adjacent fields, thereby generating a composite image of the ROI – The camera's specifications define the ROI surface of $8377 \times 6990\ pixels$, i.e. 2.29mm × 1.91mm. The BF-images were rescaled to have 0.5µm-pixel size ($4580 \times 9820\ pixels$), which is in the same order of the diffraction limit.

The acquisition procedure is illustrated in Figure 4.

### 1.2.3 Image registration

First, BF-ROI and BF-H&E-ROI were registered at a resolution of 0.5 µm per pixel. This step aimed to correct morphological distortions in the native tissue section caused by the histochemical staining process – including hydration, staining, and dehydration.

Manual registration was first conducted by applying a thin-plate spline transformation to the BF-H&E-ROI, calculated from, at least, four anatomical landmarks as reference points. The roughly aligned ROIs were subsequently used to train an automated finer registration pipeline. This pipeline utilized an eosin-style transfer model, including a color (or stain) deconvolution (Ruifrok and Johnston, 2001), and followed by a histogram matching and a sequential rigid-to-deformable registration (translation, affine and B-spline deformable transformations) of the true eosin ROI onto the one inferred from the BF-ROI, as shown in Figure 5. We emphasize that, at this stage of the image processing pipeline, the eosin-style transfer model does not achieve the image quality required for pathologist assessment.

The second registration consisted of extracting the IR-ROI in the IR-WSI at 8µm-precision (IR image pixel size) aligned with the BF-ROI. To achieve this, a BF-WSI of the unlabeled tissue section acquired at magnification 5× served as the cross-reference between the BF-ROI and the IR-ROI.

The IR-WSI was first registered onto the BF-WSI rescaled at pixel size 8µm through an affine transformation calculated from 3 landmarks manually selected. Additionally, the BF-ROI downscaled at magnification x5 was localized (correlation ROI matching) and finely aligned (similarity transformation) into the original BF-WSI. Based on these findings, the IR-ROI was easily retrieved in the IR-WSI and aligned to the BF-ROI using the inverse (similarity) transformation.

### 1.3 Deep Learning algorithm architectures

Our digital pathology model is based on a conditional GAN framework (Mirza and Osindero, 2014). Our generator ($G$) is based on an attention U-Net architecture (Oktay et al., 2018) that we modified for multimodal and multispectral input data and named Attention Multimodal UNet (AttMUNet). We refer to the overall architecture as "MUGAN". Our discriminator ($D$) is a patchGAN classifier, see Figure 6. PatchGAN works by splitting a patch into sub-patches. Then, it tries to classify each of these thumbnails as real or fake (in our case: a ground truth chemical H&E, or a digitally stained one). All of these responses are averaged to provide a single response (real or fake) for the entire patch at the output of $D$. Both generator and discriminator use modules of the form Convolution-BatchNorm-LeakyReLu (see Figure 7 and Figure 8).

The AttMUNet (Figure 7) is designed to process and integrate features from both BF image patches and IR image patch stacks. It extracts and fuses morphological and spectral features from these modalities to enable digital H&E staining. Because the BF and IR data have different spatial resolutions (0.5 µm and 8 µm, respectively), the IR data is injected later in the network than the BF data. This ensures that their feature maps have compatible dimensions after the convolution and max-pooling stages, allowing them to be concatenated and passed together to the decoder. The learning rate for optimizing $G$ was set at $1.10^{-4}$, while for $D$, it was set at $1.10^{-5}$. The Adam optimizer was employed for network training, and the batch size was set as 24.

The generator $G$ is trained to produce images that the discriminator $D$, which is simultaneously trained to distinguish real images from the generator's fakes, cannot tell apart from real ones. To force the generator to produce an output with the same structure as the BF-patch, both BF and H&E-patch are concatenated before being fed to $D$. It conditions $D$ based on the BF-patch. We define $x_{BF}$ as BF-patch input, $x_{IR}$ as IR-patch input, $x$ as the pair of inputs $(x_{BF}, x_{IR})$, $y$ as the H&E-patch reference and $\tilde{y}$ as H&E output generated by $G$. The objective of our conditional GAN can be addressed as:

$$\mathcal{L}_{cGAN}(G,D) = \mathbb{E}_{x,y}(\log D(x_{BF},y)) + \mathbb{E}_{x,y}(\log(1 - D(x_{BF},G(x))) \quad (3)$$

In equation (3), $G$ tries to minimize this objective, and $D$ tries to maximize it, meaning that our global objective $G^* = \arg\min_G \max_D \mathcal{L}_{cGAN}(G,D)$. We also add a $L_1$ (Mean Absolute Error) loss ($\mathcal{L}_{L1}$), controlled by a regularization parameter $\lambda$ (set to 100), to capture low frequency features and a $LPIPS$ (Learned Perceptual Image Patch Similarity) perceptual loss ($\mathcal{L}_{LPIPS}$) (Zhang et al., 2018) to capture high frequency features and produce realistic H&E patches. Our final global objective is now:

$$G^* = \arg\min_G \max_D \mathcal{L}_{cGAN}(G,D) + \lambda \mathcal{L}_{L1}(G(x),y) + \mathcal{L}_{LPIPS}(G(x),y) \quad (4)$$

In equation (4), $L_1$ is defined as:

$$L_1 = \frac{1}{N}\sum_{i=1}^{N}|y_i - \tilde{y}_i| \quad (5)$$

With $N$, $y_i$ and $\tilde{y}_i$ the total number of pixels, the ground truth pixel value at position $i$ and the predicted pixel value at the same position $i$, respectively.

The $LPIPS$ loss is defined as a distance between two images $x$ and $y$ with a network (in this case a Visual Geometry Group neural network – VGG). A stack of $L$ layers is extracted and unit-normalized (designated as $\hat{f}_l$) in the channel dimension for each layer $l$. The activations are scaled channel-wise by vector $w_l$ and the $L_2$ distance is then computed:

$$LPIPS(x,y) = \sum_l w_l \frac{1}{H_l W_l} \sum_{h=1,w=1}^{H_l W_l} \|\hat{f}_l(x)_{hw} - \hat{f}_l(y)_{hw}\|_2^2 \quad (6)$$

More specifically, the loss used for training the discriminator (D) was defined as:

$$\mathcal{L}_D = \frac{BCE(D(x,y)) + BCE(D(G(x),y))}{2} \quad (7)$$

In equation (7), $BCE$ is the Binary Cross Entropy (Mirza and Osindero, 2014) defined for $N$ observations:

$$BCE = -\frac{1}{N}\sum_{i=1}^{N}[z_i \log(\hat{y}_i) + (1-z_i)\log(1-\hat{z}_i)] \quad (8)$$

Where $z_i \in \{0,1\}$ is the actual binary label of the $i^{th}$ observation, and $\hat{z}_i \in (0,1)$ is the predicted probability that $z = 1$.

The discriminator should no longer distinguish the real images from the inferred ones during training, and the loss $\mathcal{L}_D$ theoretically tends to ½.

ROIs from 15 prostate tissue sections were considered: ten sections were used for training and validation, while the remaining five were reserved for testing and clinical assays.

In total:

- 131 672 triplets of patch images were extracted from 276 ROIs (18.4 ROIS per slide in average).
- 118 602 triplets were used for the training (approximately 90%) with a train/validation ratio of 90-10%.
- 13 070 triplets were used for the test (approximately 10%).

Data augmentation was done during training to reduce overfitting and improve model generalization: random horizontal and vertical flips were applied to the triplet of images, and random adjustments to brightness, contrast, and saturation (±10%) were applied only to BF and IR images. All transformations were applied with a probability of 0.5.

Representative image triplets are illustrated in Figure 9.

## 1.4 Preclinical evaluation – Methodology

Beyond conventional mathematical metrics for image quality assessment, we conducted statistical evaluations of both perceived image quality and diagnostic performance using pathologist-derived ratings.

We selected 109 *ROIs* from prostate tissue sections that were unknown to our model. We applied our model to get 109 H&E inferred *ROI* images that were compared with their H&E ground truth counterparts. We finally got two sets of 109 images *GT* and *IN* (ground truth set and inferences). *GT* and *IN* in equation (9) were randomly mixed up to have two complementary lists $L$ and $\bar{L}$ composed of both inferred and ground true images, equation (10).

$$\begin{array}{c} GT = \{GT_1, \dots, GT_{109}\} \\ IN = \{IN_1, \dots, IN_{109}\} \end{array} \quad (9)$$

$$(GT, IN) \xrightarrow{shuffle} (L, \bar{L}):$$
$$L = \{GT_1, IN_2, GT_3, \dots, IN_{109}\} \quad (10)$$
$$\bar{L} = \{IN_1, GT_2, IN_3, \dots, GT_{109}\}$$

The evaluation was then performed in two phases : first $L$ is sent to the pathologist $P_1$ since $\bar{L}$ was sent to $P_2$ for evaluation. After three weeks (the washout duration), $P_1$ evaluated $\bar{L}$ and $P_2$ evaluated $L$, see Figure 10. Pathologists rated the image quality based on criteria presented in Table 1 & Table 2, as defined by the French Association for Quality Assurance in Pathology – AfAQap (Egele, 2024). The Table 2 describes the notation (rating from 1 to 4) used in the rest of our study.

The evaluation was then completed by statistics applied on the Gleason grade set on the ground truth and inferred ROIs.

This score is used by all pathologists to assess the aggressiveness of prostatic adenocarcinoma. It is calculated by adding the grade of the most common cancer pattern to the grade of the second most common pattern. Each grade ranges between 3 and 5 and the Gleason score ranges from 6 (3+3) to 10 (5+5) (Epstein et al., 2016; JLH van Leenders et al., 2020). More the score is high, worse is the prognostic. The rating used within this study is summarized in the Table 3.

## 2 Results and discussion

### 2.1 Qualitative results

Figure 11 (A) and (B) present infrared images (2.74 cm × 2 cm) of the same unstained prostate tissue slide (#B00980; see SI), acquired at wavelengths targeting nucleic acids and proteins. These images reveal absorption contrasts corresponding to ribose (1041 cm$^{-1}$)-phosphate (1241cm$^{-1}$) vibrational modes and amide I-II (1641 cm$^{-1}$ / 1556 cm$^{-1}$) vibrational modes, respectively. The absorption band around 1040 cm$^{−1}$ reflects overlapping contributions from ribose-related RNA vibrations and carbohydrate C–O stretching. In prostate tissue, which is characterized by high epithelial cellularity and limited glycogen content, this band is expected to be predominantly influenced by RNA-associated vibrations. Distinct absorption contrasts are evident between the target and closest reference images at protein-specific wavelengths, while nucleic acid-associated contrasts, although less pronounced, remain sufficiently above noise and background levels – as demonstrated by the accompanying histograms – to be effectively used by the model. The histograms exhibit two distinct peaks: a sharp peak corresponding to the background (no tissue) and a broader peak associated with the tissue, and more specifically, with the probed biochemical bound.

Figure 12 depicts the ROI highlighted by the orange box in Figure 11. It displays IR images acquired at four key wavelengths: two targeting nucleic acids (A–B) and two targeting proteins (C–D). Diffraction artifacts, particularly at histological structure boundaries, result in blurred contours of larger features. Smaller structures, especially those below the 5 µm threshold, may exhibit reduced contrast. Additionally, diffraction-induced constructive interference can produce localized hot spots, requiring careful management of the bolometer's dynamic range.

Figure 13 and Figure 14 present comparisons between the true and digital H&E staining for two ROIs from two tissue sections (#B00980 and (#B00978). It is important to note that the ROIs were realized from the same tissue, not adjacent tissues, ensuring full comparability. From left to right, the digital H&E generated by our model and the ground truth H&E are shown. Notably, cytological structures as well as the color contrast in the digital H&E image, are faithfully reproduced and closely resemble those observed in the true H&E staining. In particular, the shape of cells, nuclei, glands, vascular structures, and their contents, as well as connective tissue, are reproduced with very high accuracy. The colorimetry is also comparable to chemical H&E staining.

### 2.2 Quantitative results – Metrics

Beyond the qualitative comparisons, we have considered four quite common metrics computed in image processing to measure image similarity including, from the simplest to the most complex in terms of computation complexity : Peak Signal-to-Noise Ratio ($PSNR$) (Oktay et al., 2018) , Structural Similarity Index Measure ($SSIM$) (Horé and Ziou, 2010), Multi-Scale Similarity Index Measure ($MS-SSIM$) (Wang et al., 2003) and $LPIPS$ (Zhang et al., 2018).

$SSIM$ and $PSNR$ are widely used in the digital pathology community. Here, we also added the MS-$SSIM$ metric that is less sensitive than $SSIM$ to pixel shifts that can occur during the acquisition and that have been not fully corrected in the registration phase – for more information, the pixel shift impact on the $SSIM$ and $MS-SSIM$ is presented in the SI. $LPIPS$ is also well suitable for a human perception based comparison. Ideally, the $SSIM$ and $MS-SSIM$ should reach a value of 1 (arrow up), $PSNR$ should tend to $+\infty$ and $LPIPS$ should tend to 0 since it corresponds to a distance between the ground truth features and the inferred counterparts (0 means the two images are considered as identical by the neuron network). ROI backgrounds were excluded from metrics computations. This avoids any bias that would lead to an artificial overestimation of the performance (in particular of the SSIM and PSNR metrics). For more information on the overall data treatment process, please refer to the SI.

Metrics computed on the test dataset lead to quantitative results : $SSIM = 0.78\ (\pm 0.057)$, $MS-SSIM = 0.82\ (\pm 0.072)$, $PSNR = 23.71\ (\pm 2.454)$ and $LPIPS = 0.039\ (\pm 0.039)$. $PSNR$ and $SSIM$ are reported in Table 4, which also compiles the results of the most representative studies from the literature applied to virtual H&E staining. Although experimental context and conditions (organ, resolution, acquisition type, test database size and test set preparation) may differ from our work, it still offers a literature baseline to compare our work. We excluded from this table any paper reporting a small number of test ROIs, applying pretreatment such as blurring or presenting results from a validation dataset, as these strongly overestimate performances. We intentionally did not include results obtained using complex spectroscopic approaches, such as photoacoustic or stimulated Raman imaging (Sarri et al., 2019; Boktor et al., 2024; Liu et al., 2024). Although these methods are scientifically compelling, in our view, their technical complexity currently limits their adoption in routine laboratory settings.

We emphasize that this comparison provides a useful overview of current performance levels but should be interpreted with caution, as the results were obtained at different magnifications (×20 and ×40), and the reported number of patches is approximate, since other studies do not necessarily compute metrics on 256×256-pixel patches. Notably, image registration, which strongly affects $SSIM$ and $PSNR$, is even more critical at ×40 magnification.

In our case, $PSNR$ and $SSIM$ exhibit pretty high values. The $PSNR$ values (typically ~23), are above values reported in the Table 4. The $SSIM$ values of approximately 0.78 are already satisfactory since the $MS-SSIM$ values, which is less pixel-shift sensitive, reach up to 0.84 (note reported in the Table 4) that is indicative of strong inference performance. These $SSIM$ and $MS-SSIM$ performances suggest that our network maintains cell/tissue architecture and structure fairly well. The autofluorescence-based approach is close to our methodology in terms of tissue handling and image preparation. The registration component in (Li et al., 2024), which is similar to the VoxelMorph framework (Balakrishnan et al., 2019), appears, in our view, to be slightly more efficient than our approach and results in a marginally higher $SSIM$.

In the brightfield-based approach described by (Rana et al., 2020), samples used for training are reused for testing, which, in our view, leads to a clear overestimation of performance. Moreover, we assume that the image background is retained, which would further inflate the reported metrics. Finally, the registration performed using Photoshop results in the $SSIM$ value that appears unexpectedly high – especially considering such a large number of images.

$LPIPS$ values obtained by our staining model are within a good range and suggest that morphological structures are likely preserved and there are only moderate perceptual differences between our inferred H&E and the ground truth. Our results are very encouraging in the context of the literature, they are better than other conditional GAN-based work, and competitive to more recent diffusion-based methods.

## 2.3 Statistical analyses – Preclinical validation

Beyond qualitative and quantitative evaluation of the digital images provided by our approach, we went a step further and conducted a preclinical evaluation led by two pathologists regarding the image quality rating and diagnosis / prognosis scores.

### 2.3.1 Statistical analyses on image quality scores

The objective of the first analysis is to determine whether our multimodal infrared digital staining can achieve the same image quality as H&E chemical staining.

The results of quality ratings carried out by the two pathologists are summarized in Figure 15 (A). At first sight, there is a good agreement between inferred image quality and image quality of ground truths. The ratings are mainly between 3 and 4 demonstrating the good overall quality of colors and morphologies of tissues reproduced by our approach. The majority of low scores (< 3) were attributed to inferred ROIs (10/12 cases), indicating some room for improvement, particularly in the rendering of nuclear components. In most ROIs, $P_1$ considered that the collagen appeared a bit too red and set a score of 3. The Table 5 gives the percentage of ROIs considered optimal (rating = 4) and good (rating≥3 in brackets) for diagnosis or prognosis for the three main cellular elements in the ground truth and inferred ROIs – see Table 1 for the selected criteria. These results show that the digital rendering of nuclear pixels can be further optimized, even when pathologists deem the current image quality sufficient for diagnostic purposes. Notice that near 100% of digitally stained ROIs received a score of 3 (good) for the three categories.

We applied the $\chi^2$ two-sided test of homogeneity (Ground truth versus inferred) (Faizi and Alvi, 2023) on the image quality scoring with null hypothesis $H_0: p_{GT} = p_{IN}$ for each cellular components (nucleus, cytoplasm and collagen) to objectively quantify the difference between inferred and ground truth ROIs. It turns out that there is rejection of homogeneity between ground truth and digital scores for nucleus for $P_1$ and for cytoplasm for $P_2$ (5%- significance level). For other elements, ground truth ROI quality and inferred ROI quality are not statistically differentiated, with a quality of collagen equivalent for both pathologists. These results suggest that our digital staining is equivalent to the gold-standard chemical H&E stain demonstrating the effectiveness and the usability of our digital staining model in pathology.

### 2.3.2 Statistical analyses on the Gleason scores

The objective of this study is to answer the question: are digitally stained slides equivalent to H&E chemically stained images for prostate cancer diagnoses and prognoses ?

The gradings of every ROIs are presented in the Figure 15 (B-C). To establish the statistic, we have also added the category "IHC" in case of the pathologist cannot set a definitive diagnosis that would require a complementary immunohistochemistry analysis. The intra-pathologist agreement is 86.2% for ground truth ROIs and 84.4% for inferred ROIs. The agreement between chemical H&E staining and our digital staining is 88% and 87.2% for $P_1$ and $P_2$ respectively (Figure 15 (D)).

We applied the $\chi^2$ two-sided test of homogeneity (Ground truth versus inferred) on the Gleason scoring with null hypothesis $H_0: p_{GT} = p_{IN}$. The statistics are presented in Table 7. It turns out that gradings done from inferred ROIs are statistically equivalent to the ones from ground truth ROIs. Diagnoses or prognoses based on observation of images digitally stained by our DL model are as meaningful as those based on chemically stained tissue images. Complementary elements that confirm the results are shown in SI.

## 3    Conclusion and perspectives

We have introduced a fast bimodal microscopy platform combining brightfield and infrared imaging, along with a deep learning-based optical staining model capable of processing both modalities. Our results demonstrate strong agreement between our digital staining method and the chemical H&E staining reference. Notably, our approach excels in accurately rendering nuclear and nucleolar structures. We have demonstrated our capability to get IR-WSI over 5 cm² tissue surface in few minutes making this approach fully compatible with pathology workflow constrains.

The automatic registration and the deep learning staining model have led to pretty good metrics on test WSI samples (over 23 for $PSNR$, and close to 0.8 for $SSIM$ and $MS-SSIM$). The performance metrics obtained with our IR-based digital staining even surpass those reported in the literature. To further benchmark our digital imaging approach against the histopathological gold standard, a preclinical validation study, conducted in collaboration with expert pathologists, has then been presented. Within this preclinical validation framework, statistical analyses based on morphological criteria – such as nuclear, cytoplasm and extra-cellular features – and Gleason grading have been provided for a rigorous and clinically meaningful assessment of our model performance.

Our IR-based digital staining achieves image quality sufficient to support equivalent diagnostic and prognostic assessments as those performed on conventionally stained slides. The image quality scoring of digitally stained ROIs has been considered good enough by pathologists, sometimes even optimal. Most of the time, the statistic has demonstrated that the scores obtained from the digital images cannot be distinguished from the images obtained from current H&E staining (the agreement between inferred and ground truth ROIs reached 88%). However, detailed analysis of ROI images – systematic observations and comparisons of images – and statistics show that nucleus rendering, in terms of number and details, can still be improved. We are currently working on enhancing IR-spatial resolution to enhance nuclei detection in particular. In addition, we are improving the architecture of the DL model by replacing the CNN by a visual transformer architecture for the Generator. The DL model developed for prostate tissues serves as a basic model that we are generalizing to colon, breast and skin tissues. The infrared-derived features enhance this task by introducing novel metabolomic insights not captured in prior approaches. This model is even expected to be generalized to any tissue types. The IR-setup presented in this paper is also being embedded into a bimodal IR-VIS-scanner with some modification on the IR-optical path to gain in compactness. A first prototype is shown in the Figure 1.

Looking forward, this infrared-driven optical staining approach will be further advanced through the integration of metabolomic maps – such as collagens, lipids, carbohydrates, and selected proliferation markers – overlaid onto H&E-like images (Kallenbach-Thieltges et al., 2020; Gerwert et al., 2023). Ultimately, this work lays the foundation for broader adoption of infrared spectroscopy in pathology laboratories, facilitating the identification of advanced diagnostic and prognostic markers based on complex IR molecular signatures that will be superimposed to our digital H&E images.

# 4 Reference

Amrania, H., Woodley-Barker, L., Goddard, K., Rosales, B., Shousha, S., Thomas, G., et al. (2018). Mid-infrared imaging in breast cancer tissue: an objective measure of grading breast cancer biopsies. *Converg Sci Phys Oncol* 4, 025001. doi: 10.1088/2057-1739/aaabc3

Bai, B., Wang, H., Li, Y., de Haan, K., Colonnese, F., Wan, Y., et al. (2022). Label-Free Virtual HER2 Immunohistochemical Staining of Breast Tissue using Deep Learning. *BME Front* 2022. doi: 10.34133/2022/9786242

Bai, B., Yang, X., Li, Y., Zhang, Y., Pillar, N., and Ozcan, A. (2023). Deep learning-enabled virtual histological staining of biological samples. *Light Sci Appl* 12. doi: 10.1038/s41377-023-01104-7

Baker, M. J., Trevisan, J., Bassan, P., Bhargava, R., Butler, H. J., Dorling, K. M., et al. (2014). Using Fourier transform IR spectroscopy to analyze biological materials. *Nat Protoc* 9, 1771–1791. doi: 10.1038/nprot.2014.110

Balakrishnan, G., Zhao, A., Sabuncu, M. R., Guttag, J., and Dalca, A. V. (2019). VoxelMorph: A Learning Framework for Deformable Medical Image Registration. *IEEE Trans Med Imaging* 38, 1788–1800. doi: 10.1109/TMI.2019.2897538

Balan, V., Mihai, C. T., Cojocaru, F. D., Uritu, C. M., Dodi, G., Botezat, D., et al. (2019). Vibrational spectroscopy fingerprinting in medicine: From molecular to clinical practice. *Materials* 12. doi: 10.3390/ma12182884

Barth, A. (2007). Infrared spectroscopy of proteins. *Biochim Biophys Acta Bioenerg* 1767, 1073–1101. doi: 10.1016/j.bbabio.2007.06.004

Beć, K. B., Grabska, J., and Huck, C. W. (2020). Biomolecular and bioanalytical applications of infrared spectroscopy – A review. *Anal Chim Acta* 1133, 150–177. doi: 10.1016/j.aca.2020.04.015

Bird, B., and Baker, M. J. (2015). Quantum Cascade Lasers in Biomedical Infrared Imaging. *Trends Biotechnol* 33, 557–558. doi: 10.1016/j.tibtech.2015.07.003

Boktor, M., Tweel, J. E. D., Ecclestone, B. R., Ye, J. A., Fieguth, P., and Haji Reza, P. (2024). Multi-channel feature extraction for virtual histological staining of photon absorption remote sensing images. *Sci Rep* 14. doi: 10.1038/s41598-024-52588-1

Cheol Park, S., Kyu Park, M., and Gi Kang, M. (2003). Super-resolution image reconstruction: a technical overview. *IEEE Signal Process Mag* 20, 21–36.

Cotran, R. S. ., Kumar, Vinay., Collins, Tucker., and Robbins, S. L. . (1999). *Robbins pathologic basis of disease*. Saunders.

Coutard, J. G., Brun, M., Fournier, M., Lartigue, O., Fedeli, F., Maisons, G., et al. (2020). Volume Fabrication of Quantum Cascade Lasers on 200 mm-CMOS pilot line. *Sci Rep* 10. doi: 10.1038/s41598-020-63106-4

Egele, C. (2024). R_RA_EN033_Rapport Technique de préparation histologique et Coloration HE 2024.

Ellis, B. G., Ingham, J., Whitley, C. A., Al Jedani, S., Gunning, P. J., Gardner, P., et al. (2023). Metric-based analysis of FTIR data to discriminate tissue types in oral cancer. *Analyst* 148, 1948–1953. doi: 10.1039/d3an00258f

Epstein, J. I., Zelefsky, M. J., Sjoberg, D. D., Nelson, J. B., Egevad, L., Magi-Galluzzi, C., et al. (2016). A Contemporary Prostate Cancer Grading System: A Validated Alternative to the Gleason Score. *Eur Urol* 69, 428–435. doi: 10.1016/j.eururo.2015.06.046

Faist, J., Capasso, F., Sivco, D. L., Sirtori, C., Hutchinson, A. L., and Cho, A. Y. (1994). Quantum Cascade Laser. *Science (1979)* 264, 553–556. Available at: http://science.sciencemag.org/

Faizi, N., and Alvi, Y. (2023). *Biostatistics Manual for Health Research : A practical Guide to Data Analysis.*, ed. S. Masucci. San Diego: Elsevier. doi: 10.1016/b978-0-12-817491-3.00014-3

Finlayson, D., Rinaldi, C., and Baker, M. J. (2019). Is Infrared Spectroscopy Ready for the Clinic? *Anal Chem* 91, 12117–12128. doi: 10.1021/acs.analchem.9b02280

Gerdes, M. J., Sood, A., Sevinsky, C., Pris, A. D., Zavodszky, M. I., and Ginty, F. (2014). Emerging understanding of multiscale tumor heterogeneity. *Front Oncol* 4, 1–12. doi: 10.3389/fonc.2014.00366

Gerwert, K., Schörner, S., Großerueschkamp, F., Kraeft, A. –L, Schuhmacher, D., Sternemann, C., et al. (2023). Fast and label-free automated detection of microsatellite status in early colon cancer using artificial intelligence integrated infrared imaging. *Eur J Cancer* 182, 122–131. doi: 10.1016/j.ejca.2022.12.026

Hackshaw, K. V., Miller, J. S., Aykas, D. P., and Rodriguez-Saona, L. (2020). Vibrational spectroscopy for identification of metabolites in biologic samples. *Molecules* 25. doi: 10.3390/molecules25204725

Hanna, M. G., and Ardon, O. (2023). Digital pathology systems enabling quality patient care. *Genes Chromosomes Cancer* 62, 685–697. doi: 10.1002/gcc.23192

Horé, A., and Ziou, D. (2010). Image quality metrics: PSNR vs. SSIM., in *Proceedings - International Conference on Pattern Recognition*, 2366–2369. doi: 10.1109/ICPR.2010.579

Hughes, C., and Baker, M. J. (2016). Can mid-infrared biomedical spectroscopy of cells, fluids and tissue aid improvements in cancer survival? A patient paradigm. *Analyst* 141, 467–475. doi: 10.1039/c5an01858g

JLH van Leenders, G., van der Kwast, T. H., Grignon, D. J., Evans, A. J., Kristiansen, G., Kweldam, C. F., et al. (2020). The 2019 International Society of Urological Pathology (ISUP) Consensus Conference on Grading of Prostatic Carcinoma. *Am J Surg Pathol* 44, e87–e99. Available at: www.ajsp.com

Kallenbach-Thieltges, A., Großerueschkamp, F., Jütte, H., Kuepper, C., Reinacher-Schick, A., Tannapfel, A., et al. (2020). Label-free, automated classification of microsatellite status in colorectal cancer by infrared imaging. *Sci Rep* 10. doi: 10.1038/s41598-020-67052-z

Kashyap, A., Rapsomaniki, M. A., Barros, V., Fomitcheva-Khartchenko, A., Martinelli, A. L., Rodriguez, A. F., et al. (2022). Quantification of tumor heterogeneity: from data acquisition to metric generation. *Trends Biotechnol* 40, 647–676. doi: 10.1016/j.tibtech.2021.11.006

Koivukoski, S., Khan, U., Ruusuvuori, P., and Latonen, L. (2023). Unstained Tissue Imaging and Virtual Hematoxylin and Eosin Staining of Histologic Whole Slide Images. *Laboratory Investigation* 103. doi: 10.1016/j.labinv.2023.100070

Kröger, N., Egl, A., Engel, M., Gretz, N., Haase, K., Herpich, I., et al. (2014). Quantum cascade laser–based hyperspectral imaging of biological tissue. *J Biomed Opt* 19, 111607. doi: 10.1117/1.jbo.19.11.111607

Latonen, L., Koivukoski, S., Khan, U., and Ruusuvuori, P. (2024). Virtual staining for histology by deep learning. *Trends Biotechnol* 42, 1177–1191. doi: 10.1016/j.tibtech.2024.02.009

Li, Y., Pillar, N., Li, J., Liu, T., Wu, D., Sun, S., et al. (2024). Virtual histological staining of unlabeled autopsy tissue. *Nat Commun* 15. doi: 10.1038/s41467-024-46077-2

Liu, Z., Chen, L., Cheng, H., Ao, J., Xiong, J., Liu, X., et al. (2024). Virtual formalin-fixed and paraffin-embedded staining of fresh brain tissue via stimulated Raman CycleGAN model. Available at: https://www.science.org

Lorenz-Fonfria, V. A. (2020). Infrared Difference Spectroscopy of Proteins: From Bands to Bonds. *Chem Rev* 120, 3466–3576. doi: 10.1021/acs.chemrev.9b00449

Mathieu, G., Dupoy, M., Bonnet, S., Rebuffel, V., Coll, J.-L., and Henry, M. (2021). Mid infrared multispectral imaging for tumorous tissue detection., (SPIE-Intl Soc Optical Eng), 5. doi: 10.1117/12.2577221

Mirza, M., and Osindero, S. (2014). Conditional Generative Adversarial Nets. Available at: http://arxiv.org/abs/1411.1784

Niklaus, F., Vieider, C., and Jakobsen, H. (2007). MEMS-based uncooled infrared bolometer arrays: a review., in *MEMS/MOEMS Technologies and Applications III*, (SPIE), 68360D. doi: 10.1117/12.755128

Oktay, O., Schlemper, J., Le Folgoc, L., Lee, M., Heinrich, M., Misawa, K., et al. (2018). Attention U-Net: Learning Where to Look for the Pancreas., in *Conference on Medical Imaging with Deep Learning*, (Amsterdam).

Rana, A., Lowe, A., Lithgow, M., Horback, K., Janovitz, T., Da Silva, A., et al. (2020). Use of Deep Learning to Develop and Analyze Computational Hematoxylin and Eosin Staining of Prostate Core Biopsy Images for Tumor Diagnosis. *JAMA Netw Open* 3. doi: 10.1001/jamanetworkopen.2020.5111

Rivenson, Y., Liu, T., Wei, Z., Zhang, Y., de Haan, K., and Ozcan, A. (2019). PhaseStain: the digital staining of label-free quantitative phase microscopy images using deep learning. *Light Sci Appl* 8. doi: 10.1038/s41377-019-0129-y

Ruifrok, A. C., and Johnston, D. A. (2001). Quantification of histochemical staining by color deconvolution. *Anal Quant Cytol Histol* 23, 291–299.

Sarri, B., Poizat, F., Heuke, S., Wojak, J., Franchi, F., Caillol, F., et al. (2019). Stimulated Raman histology: one to one comparison with standard hematoxylin and eosin staining. *Biomed Opt Express* 10, 5378. doi: 10.1364/boe.10.005378

Schnell, M., Mittal, S., Falahkheirkhah, K., Mittal, A., Yeh, K., Kenkel, S., et al. (2020). All-digital histopathology by IR optical hybrid microscopy. *PNAS* 117, 3388–3396. doi: 10.1073/pnas.1912400117/-/DCSupplemental

Senevirathna, K., Jayawardana, N. U., Jayasinghe, R. D., Seneviratne, B., and Perera, A. U. (2021). Diagnostic Value of FTIR Spectroscopy, Metabolomic Screening and Molecular Genetics in Saliva for Early Detection of Oral Squamous Cell Carcinoma (OSCC). *Medical & Clinical Research Med Clin Res* 6, 435–454. Available at: www.medclinres.org

Shi, L., Liu, X., Shi, L., Stinson, H. T., Rowlette, J., Kahl, L. J., et al. (2020). Mid-infrared metabolic imaging with vibrational probes. *Nat Methods* 17, 844–851. doi: 10.1038/s41592-020-0883-z

Sirintrapun, S. J. (2025). Review and Commentary on Digital Pathology and Artificial Intelligence in Pathology. *JCO Clin Cancer Inform*, e2500017. doi: 10.1200/cci-25-00017

Spott, A., Peters, J., Davenport, M. L., Stanton, E. J., Merritt, C. D., Bewley, W. W., et al. (2016). Quantum cascade laser on silicon. *Optica* 3, 545. doi: 10.1364/optica.3.000545

Szulczewski, J. M., Yesilkoy, F., Ulland, T. K., Bartels, R., Millis, B. A., Boppart, S. A., et al. (2024). To label or not the need for validation in label-free imaging. *J Biomed Opt* 29, 227171–227178.

Tweel, J. E. D., Ecclestone, B. R., Boktor, M., Dinakaran, D., Mackey, J. R., and Reza, P. H. (2024). Automated Whole Slide Imaging for Label-Free Histology Using Photon Absorption Remote Sensing Microscopy. *IEEE Trans Biomed Eng* 71, 1901–1912. doi: 10.1109/TBME.2024.3355296

Vitiello, M. S., Scalari, G., Williams, B., and De Natale, P. (2015). Quantum cascade lasers: 20 years of challenges. *Opt Express* 23, 5167. doi: 10.1364/oe.23.005167

Wang, Y., Guan, N., Li, J., and Wang, X. (2024). A Virtual Staining Method Based on Self-Supervised GAN for Fourier Ptychographic Microscopy Colorful Imaging. *Applied Sciences (Switzerland)* 14. doi: 10.3390/app14041662

Wang, Z., Simoncelli, E. P., and Bovik, A. C. (2003). MULTI-SCALE STRUCTURAL SIMILARITY FOR IMAGE QUALITY ASSESSMENT., in *The Thrity-Seventh Asilomar Conference on Signals, Systems & Computers*, ed. IEEE (Pacific Grove, CA, USA), 1398–1402.

Xia, Q., Yin, J., Guo, Z., and Cheng, J. X. (2022). Mid-Infrared Photothermal Microscopy: Principle, Instrumentation, and Applications. *Journal of Physical Chemistry B* 126, 8597–8613. doi: 10.1021/acs.jpcb.2c05827

Yilmaz, A., Aydin, T., and Varol, R. (2023). Virtual staining for pixel-wise and quantitative analysis of single cell images. *Sci Rep* 13. doi: 10.1038/s41598-023-45150-y

Yon, J. J., Dumont, G., Goudon, V., Becker, S., Arnaud, A., Cortial, S., et al. (2014). Latest improvements in microbolometer thin film packaging: paving the way for low-cost consumer applications., in *Infrared Technology and Applications XL*, (SPIE), 90701N.

Zhang, D., Li, C., Zhang, C., Slipchenko, M. N., Eakins, G., and Cheng, J. X. (2016). Depth-resolved mid-infrared photothermal imaging of living cells and organisms with submicrometer spatial resolution. *Sci Adv* 2. doi: 10.1126/sciadv.1600521

Zhang, R., Isola, P., Efros, A. A., Shechtman, E., and Wang, O. (2018). The Unreasonable Effectiveness of Deep Features as a Perceptual Metric., in *Conference on Computer Vision and Pattern Recognition*, (alt Lake City, UT, USA: IEEE), 586–595. Available at: http://arxiv.org/abs/1801.03924

Zhang, Y., Huang, L., Pillar, N., Li, Y., Chen, H., and Ozcan, A. (2025). Pixel super-resolved virtual staining of label-free tissue using diffusion models. *Nature Communications* 16. doi: 10.1038/s41467-025-60387-z

Zhizhina, G. P., and Oieinik, E. F. (1972). Infrared Spectroscopy of Nucleic Acids. *Russian Chemical Reviews* 41, 258–280.

**Disclosures.** L. Duraffourg, H. Borges, M. Fernandes, F. Staroz and M. Dupoy have a pending patent application related to the work reported in the manuscript.

**Acknowledgment.** L. Duraffourg et al. are grateful to Dr. Philippe Chalabreysse, M.D., for his material support and medical advice. This work benefited from the technical facilities and expertise available at Cypath and Cypath-RB, Lyon, FRANCE.

**Data availability.** Due to privacy considerations, the data supporting the findings of this study are not publicly available but may be obtained from the authors upon reasonable request.

**Conflict of Interest.** The authors declare that the research was conducted in the absence of any commercial or financial relationships that could be construed as a potential conflict of interest.

**Author Contributions.** L. Duraffourg supervised the whole project, contributed to data interpretation and wrote the manuscript; H. Borges, and M. Fernandes developed the machine learning algorithms, contributed to analysis methodology and contributed to the data interpretation; M. Beurrier-Bousquet participated to the image acquisitions and developed the bright field optical setup; J. Baraillon, B. Taurel participated to the image acquisitions and developed the infrared setup; J. Le Galudec contributed to the analysis methodology; K. Vianey, C. Maisin participated to the biological sample preparation; L. Samaison and F. Staroz contributed to methodology development and realized the medical interpretation; M. Dupoy contributed to analysis methodology, designed the study and supervised the project. All co-authors reviewed the manuscript.

**Funding.** The study was partially funded by Plan France 2030 Auvergne-Rhône-Alpes region through the "Collaborative Projects I-Demo" (MIRFLECT N°DOS0267087/00) and by the European Commission through the HORIZON-KDT-JU-2023-2-RIA project (ATHENA N°101139941).

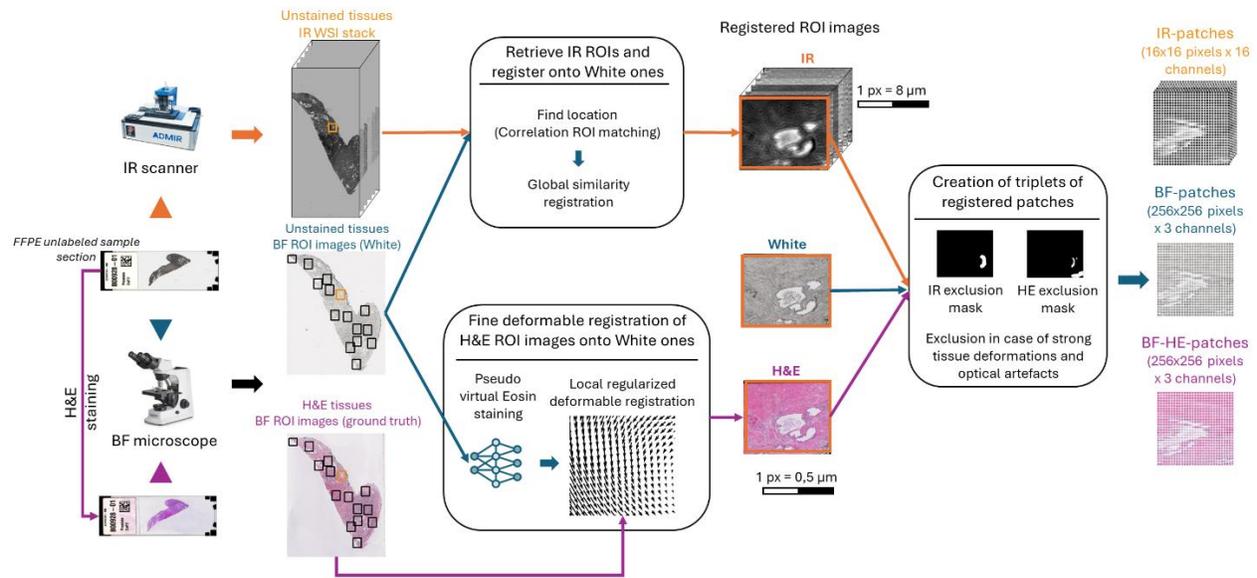

Figure 1. Synopsis of the complete method to prepare the image set for training the DL model – We automatically scan a ROI of a unlabeled tissue section with the BF-microscope. We scan the same tissue section with the IR-scanner at 16 wavelengths to build the IR WSI stack – Once these acquisitions done, this section is H&E stained with a chemical stainer and we scan the same ROIs with the BF-microscope – The BF-ROIs, both pre- and post-H&E staining, are aligned via an automated registration pipeline. This process integrates an eosin-style transfer deep learning model, histogram matching, and rigid-to-deformable image registration – IR-ROIs are extracted from the IR-WSI stack using a custom algorithm that retrieves their positions within the IR WSIs. The resulting IR unstained ROI stacks are then registered to their corresponding BF counterparts. This yields a fully aligned triplet for each ROI. The process is repeated for all ROIs across every tissue section processed by our system.

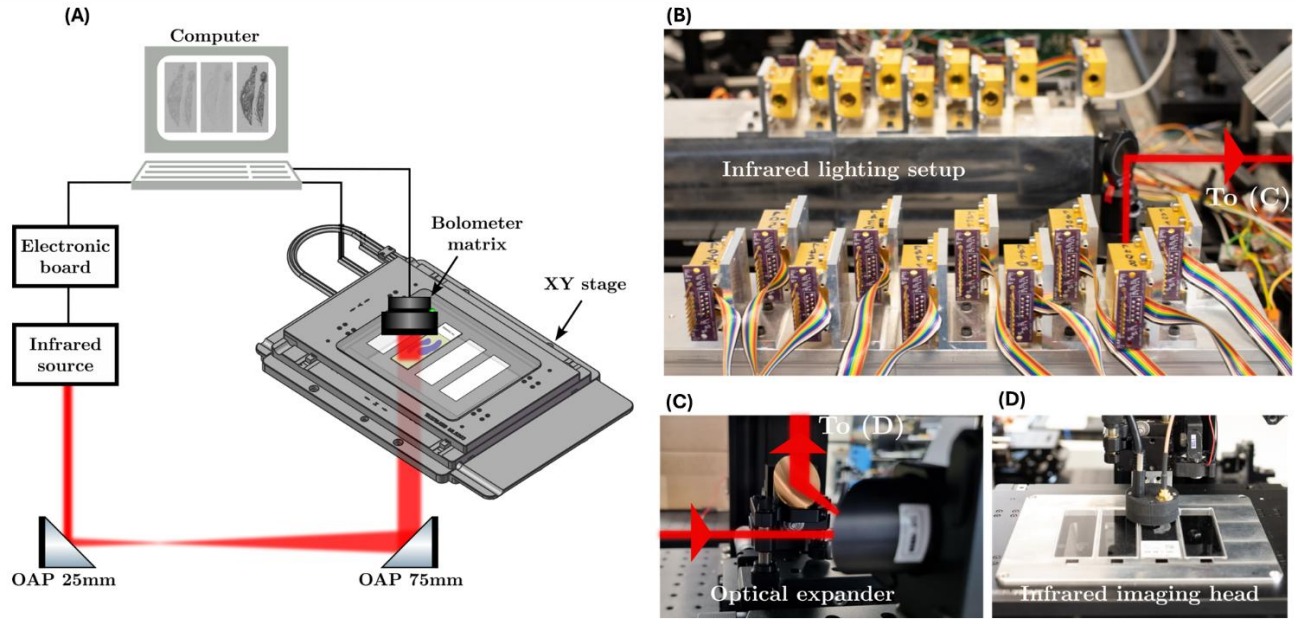

Figure 2. Infrared optical setup : (A) Global system including the imaging stage and the lighting setup – (B) Quantum Cascade laser sets forming the illumination stage – (C) Beam expander for a homogenous lighting – (D) Sample holder and IR bolometer camera.

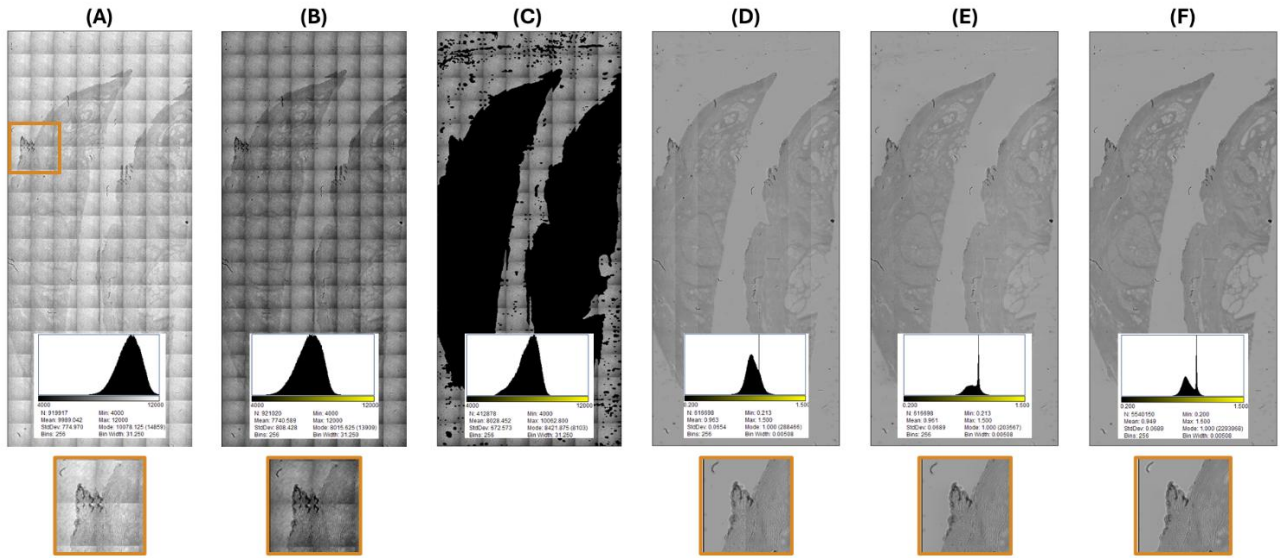

Figure 3. IR-WSI post-processing : (A) Raw image built on successive tiles corresponding to the IR camera field of view $I_{raw}$ – (B) Offset correction applied on each tile of the image $I_{offset}$ – (C) Mask definition to remove dusts and artifacts $I_{background}$ – (D) Resulting transmission WSI from equation (1) $T_{LR}$ – (E) Gradient leveling using overlaps to erase the tiling effect $T_{GL}$ – (F) Oversampling using subpixel shift acquisition to increase the spatial resolution and get 11µm-equivalent pixel size $T_{HL}$.

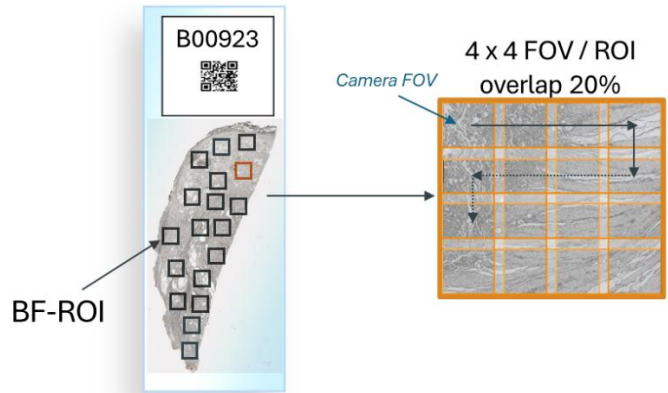

Figure 4. Acquisition method of ROIs from prostate tissue sections

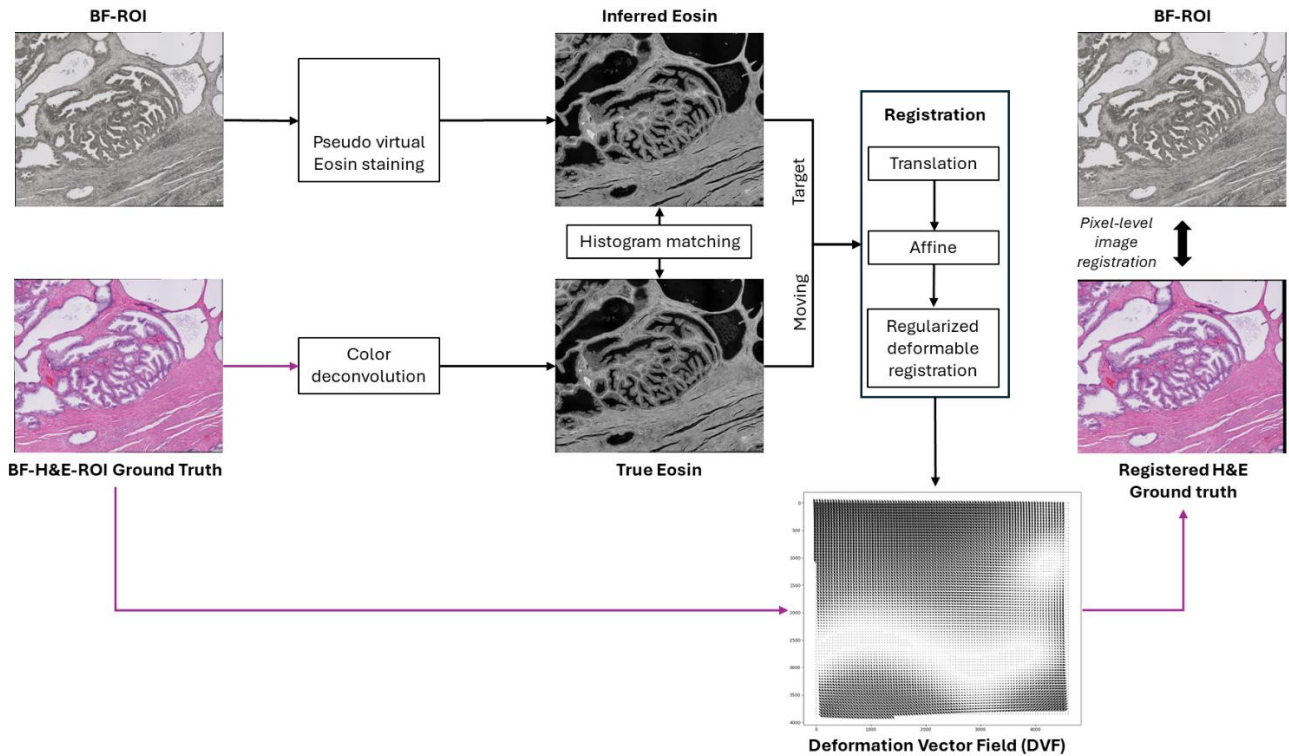

Figure 5. Registration operations between a BF-ROI and the Ground truth BF H&E ROI – The eosin-style transfer model, when applied to brightfield ROIs (BF-ROIs), preserves cytoplasmic eosin information that faithfully represents tissue contours, ensuring fine registration between the inferred eosin ROI and the true eosin ROI.

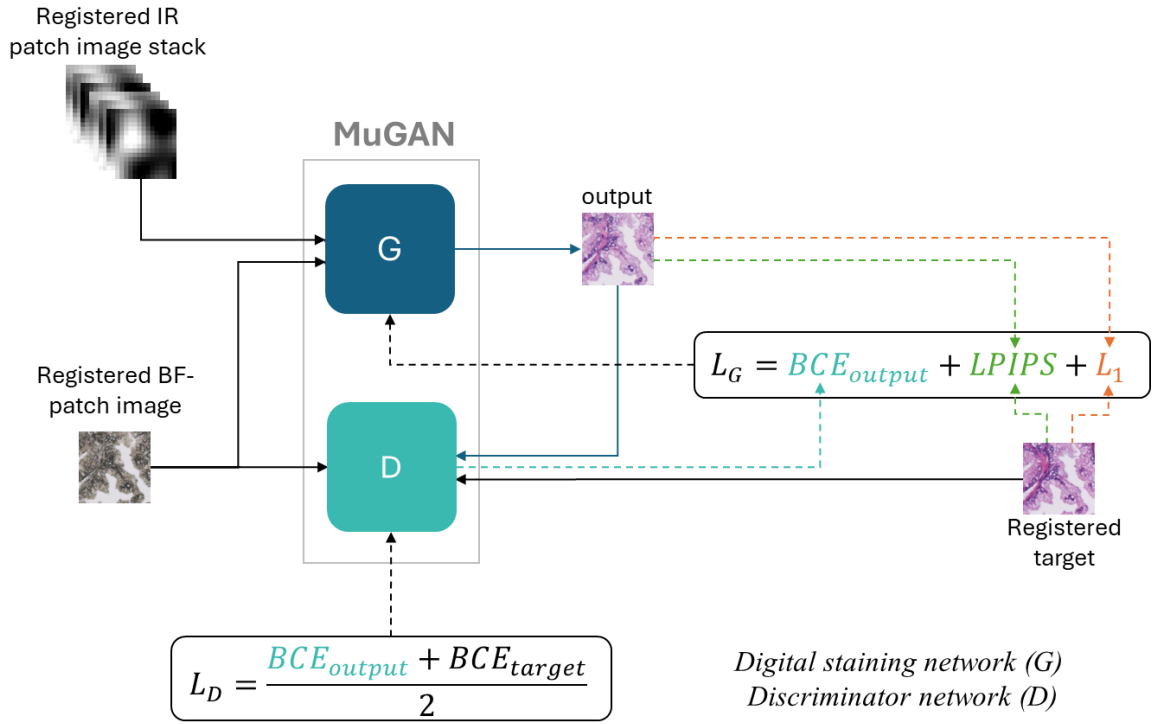

Figure 6. MUGAN framework composed of an AttMUNet generator (G) and a PatchGAN discriminator (D)

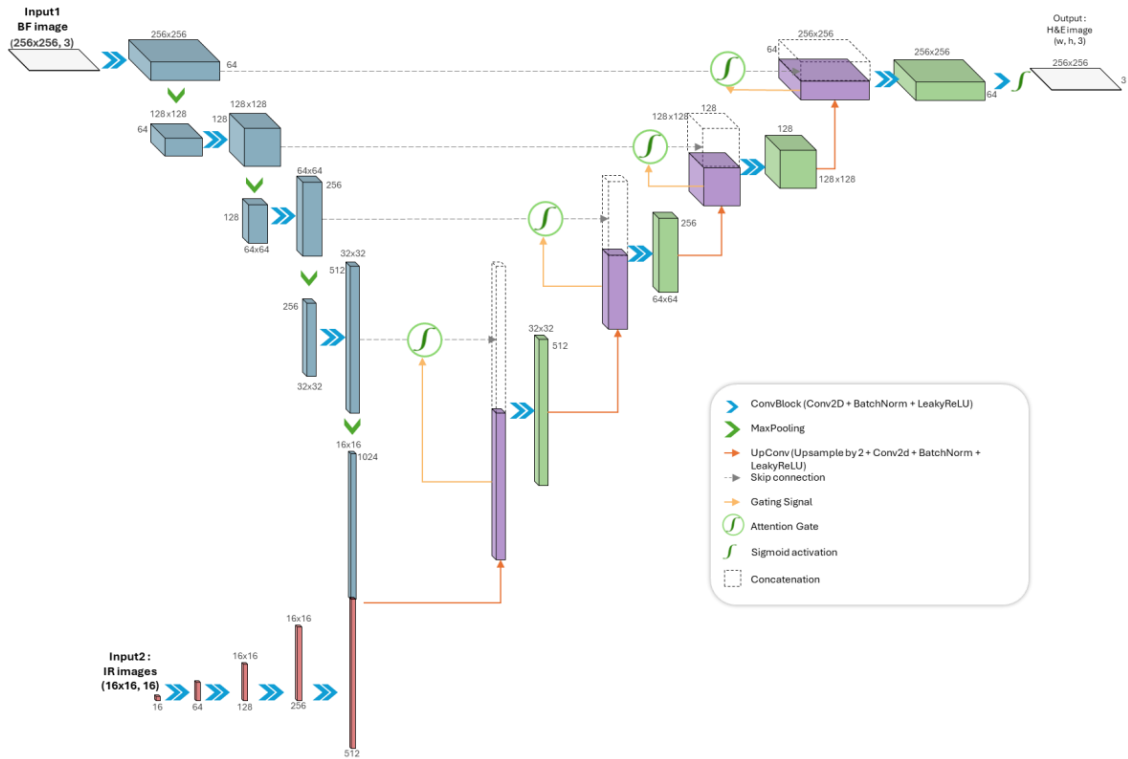

Figure 7. Schematic of the generator (G) based on the AttMUNet

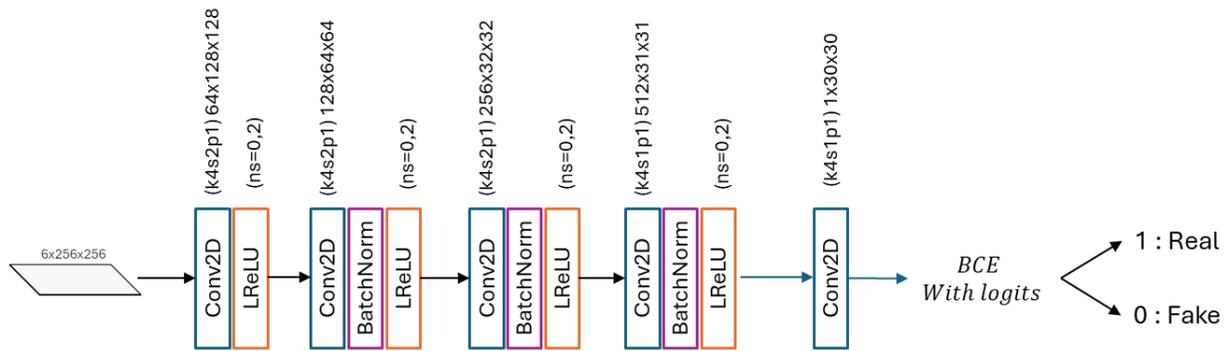

Figure 8. Schematic discriminator (D) PatchGAN – BCE : Binary Cross-Entropy – $BCE = 1 \leftrightarrow real\ image, BCE = 0 \leftrightarrow Fake\ image$

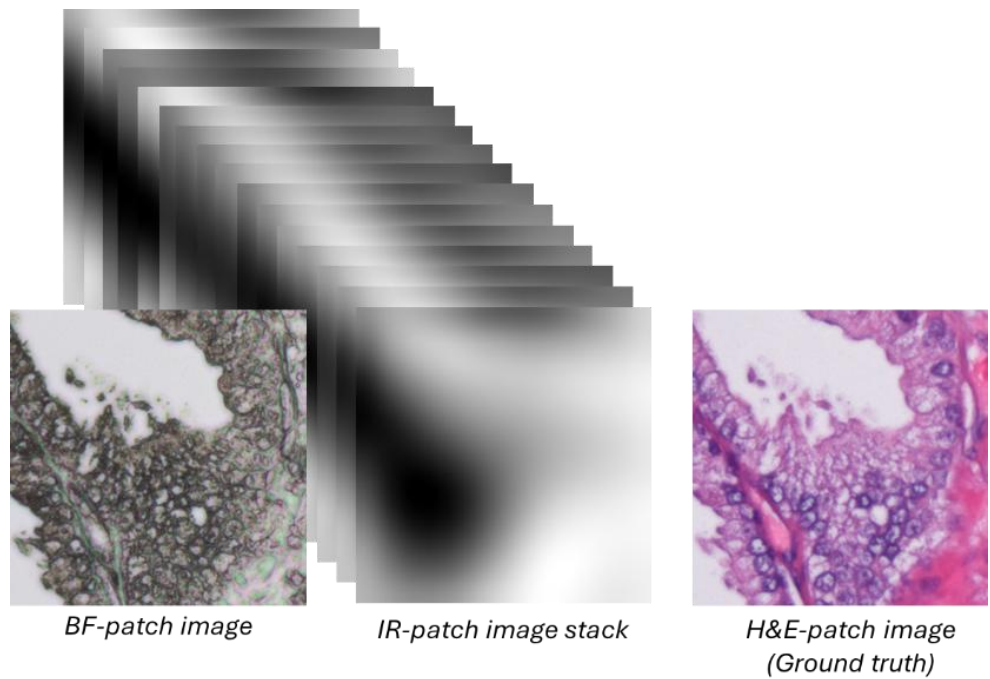

Figure 9. Typical triplet of patch images used for the training: from left to right : (256×256pixels) bright field image of unlabeled tissue, (16p×16px) IR-image stack of unlabeled tissue and (256p×256px) bright field image of H&E stained tissue. The IR images are presented in grey scale: black areas indicating large absorption while the white areas indicate low absorption.

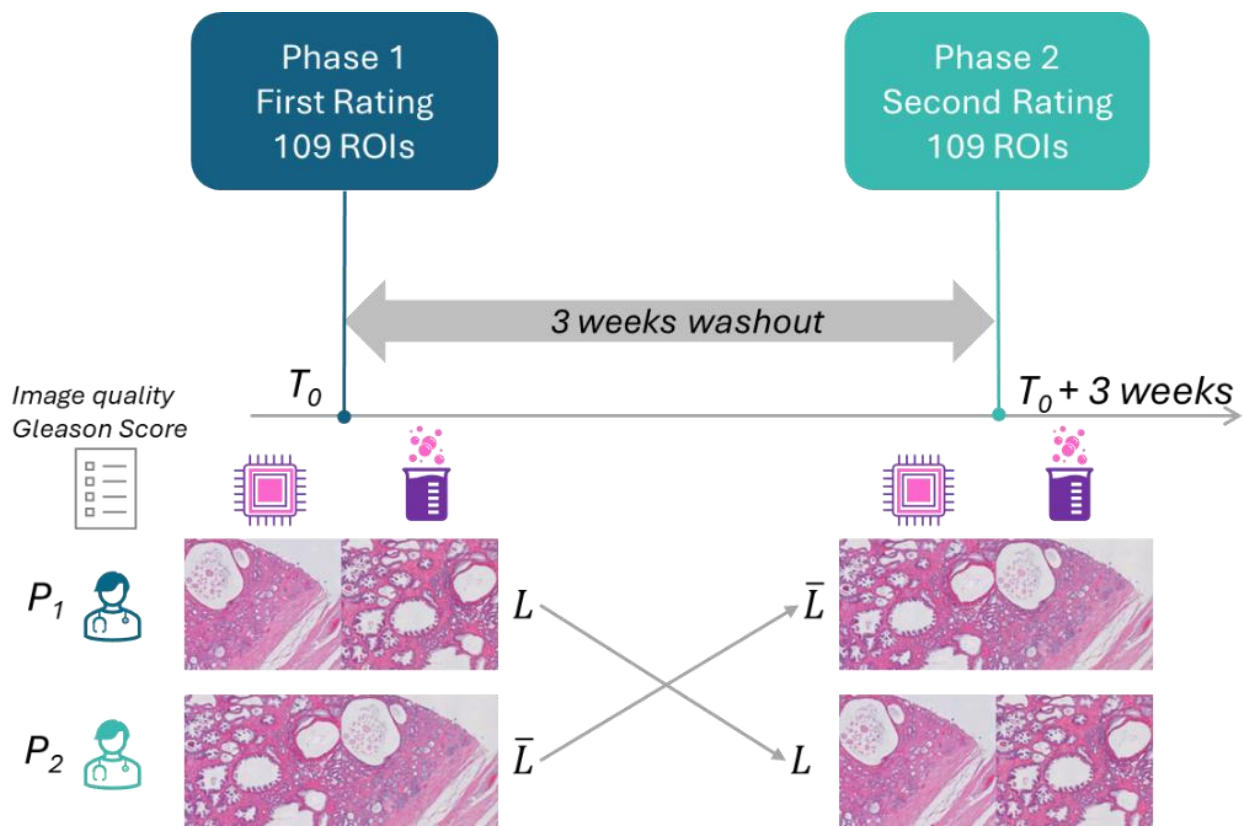

Figure 10. Preclinical assays method: We selected 109 ROIs from prostate tissue sections under test. To each ROI, it corresponds a pair of images (inferred image, ground truth) and we got two sets of 109 images (ground truth set and inferences). These two sets were randomly mixed up to have two complementary lists $L$ and $\bar{L}$ composed of both inferred and ground true images. The two sets of images were sent to two pathologists (P1 and P2) for rating (image quality rating and Gleason scores). After three weeks of washout, the two lists were swapped between P1 and P2 for a second rating.

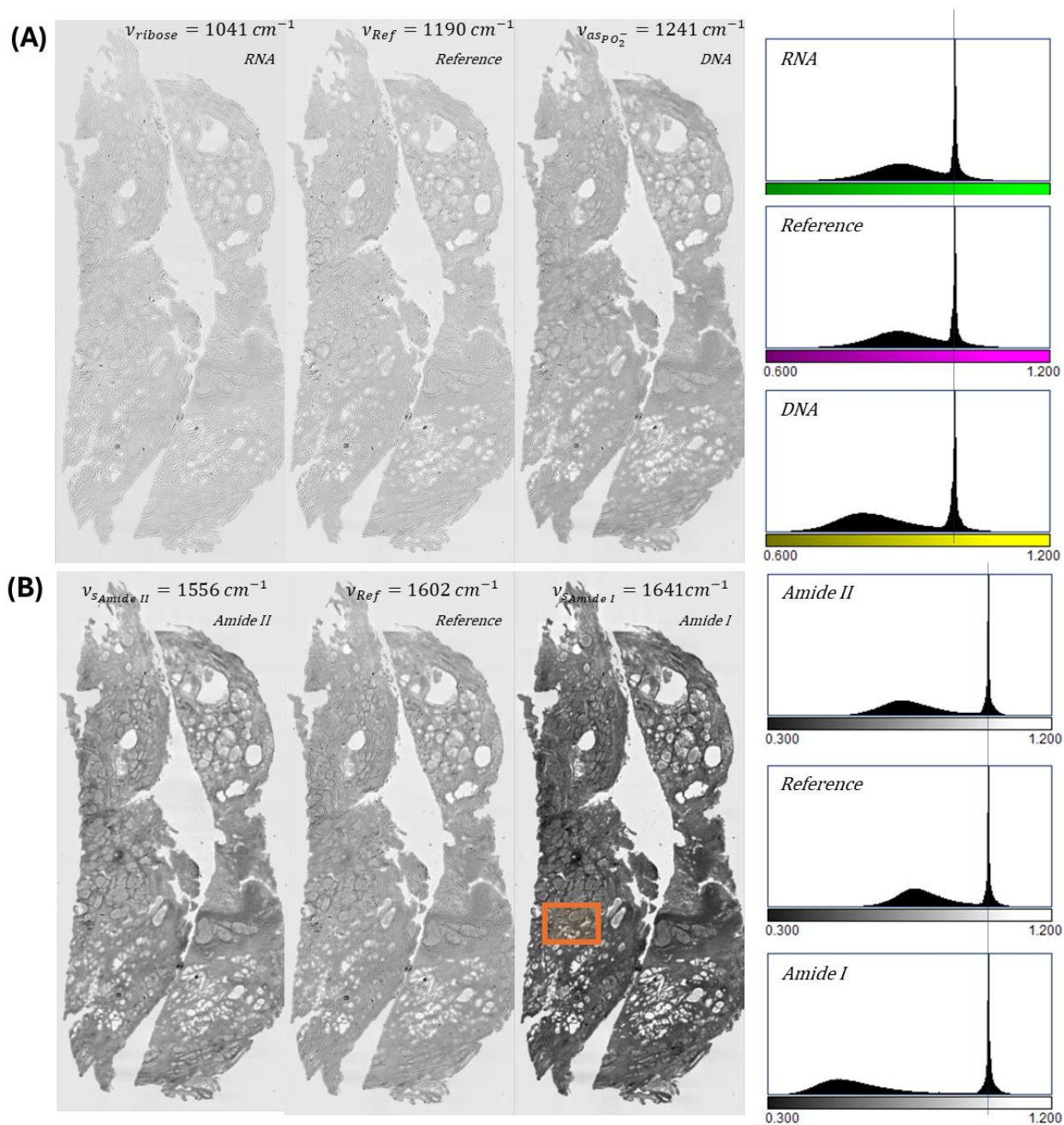

Figure 11. Whole slide images of a prostate tissue (2.74x2 cm²) (#B00980 – see SI) – (A) Infrared transmission of vibrational modes in nucleic acid, (B) Infrared image of amide vibrational modes in proteins: amide I and amid II – The corresponding histograms for each image are displayed on the right. – The orange box area is the ROI shown in the next figure.

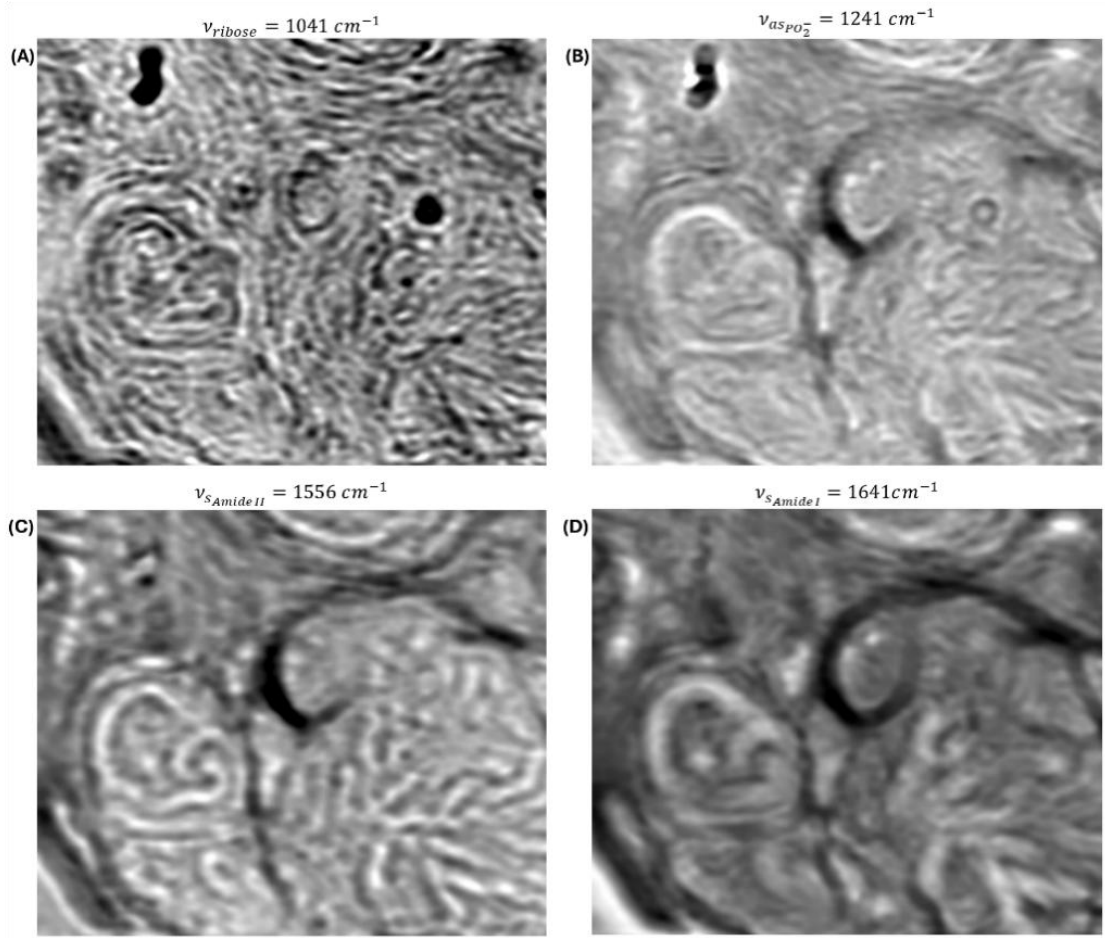

Figure 12. Images of the ROI delineated by the orange box in the previous figure (#B00980) – (A) – (D) From left to right : IR images of ribose in RNA, diphosphate in DNA, Amide II in proteins and Amide I (Helicoidal) in proteins

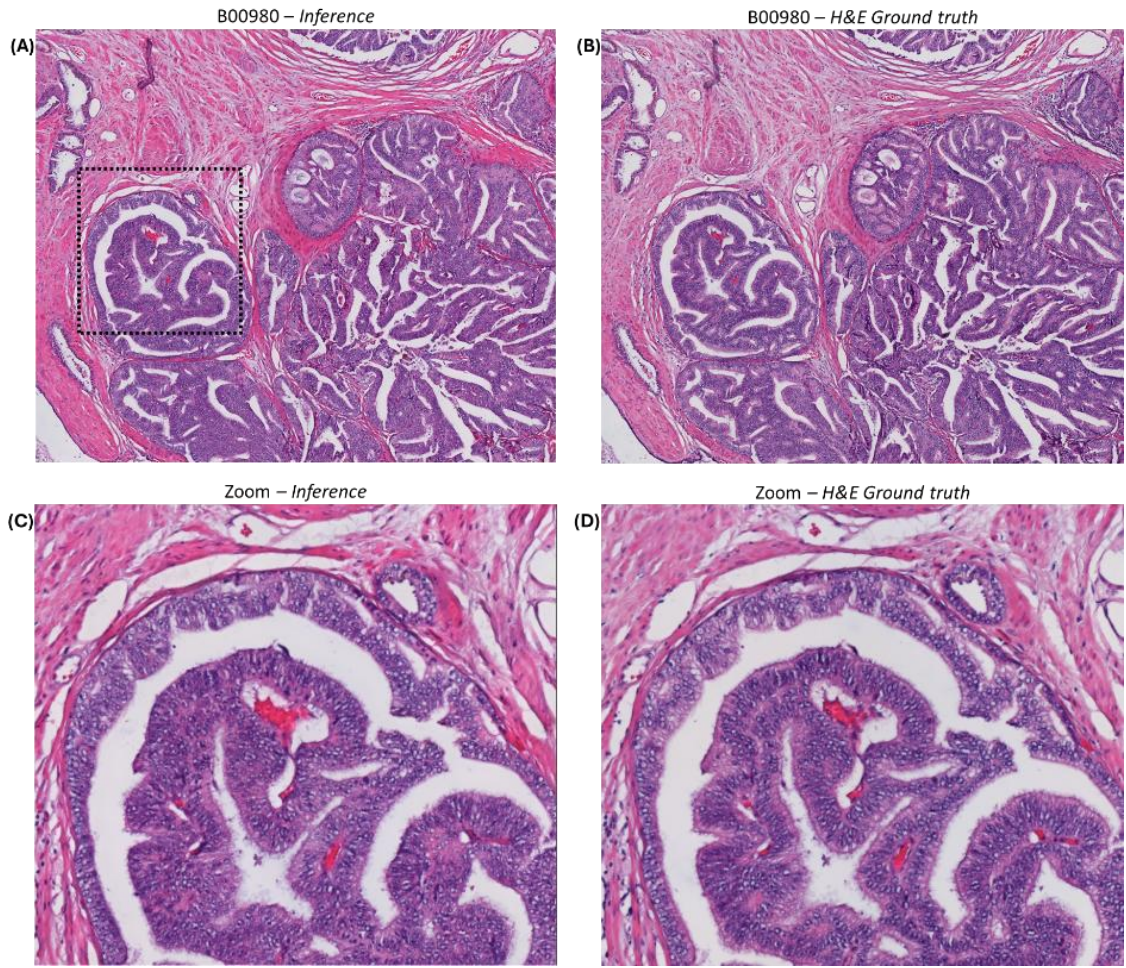

Figure 13. (A) Digital H&E inferred with our deep learning model of a ROI (#B00980) – (B) Current H&E staining of the same ROI – Ground truth – (C) Zoom on a gland area of the inferred ROI (dotted box) – (D) Zoom of the same area of the H&E stained tissue

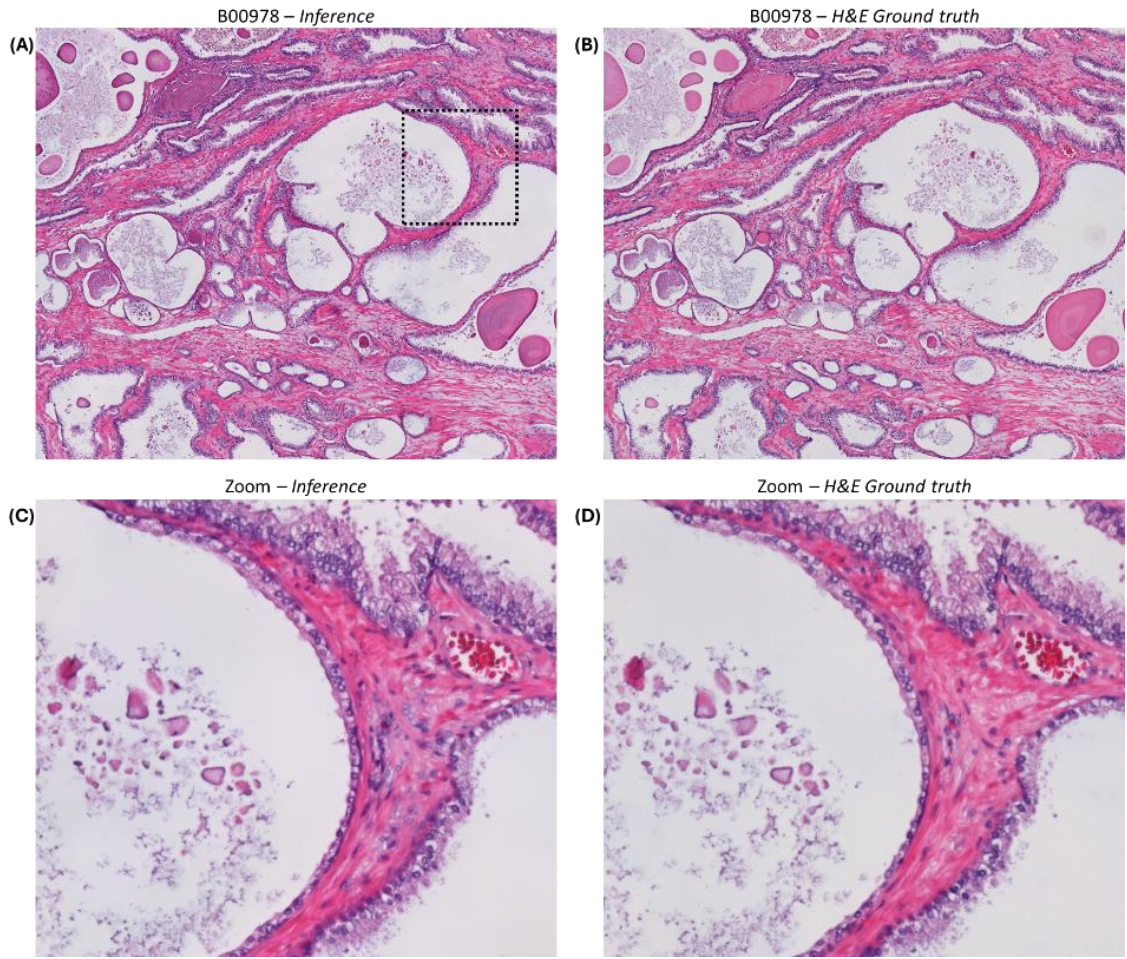

Figure 14. (A) Digital H&E inferred with our deep learning model of a ROI (#B00978) – (B) Current H&E staining of the same ROI – Ground truth – (C) Zoom on a gland area of the inferred ROI (dotted box) – (D) Zoom of the same area of the H&E stained tissue

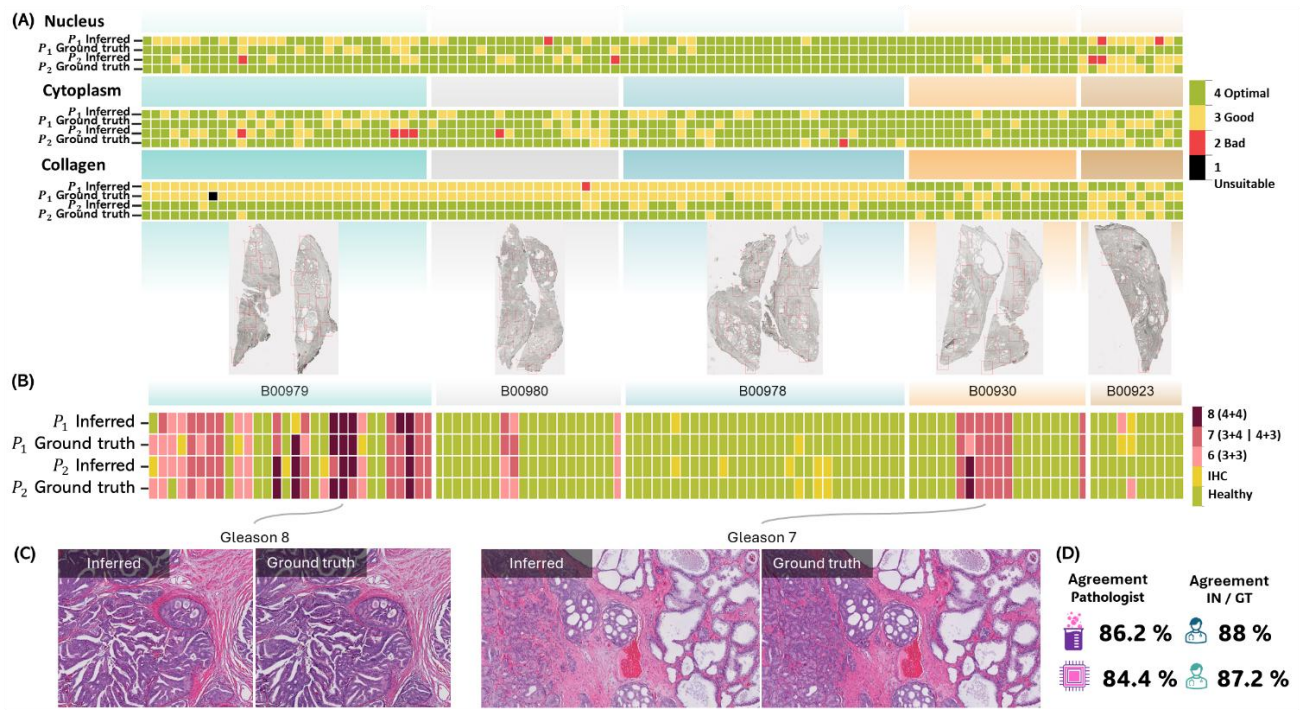

Figure 15. (A) Rating of the image quality done by the two pathologists including components: nucleus, cytoplasm and extracellular matrix (i.e. collagen), see Table 1 – Each of square column corresponds to a ROI – (B) Grading set by pathologists for the 109 ROIs (in their inferred and Ground truth versions): Comparison according to ROI – each column of square corresponds to a ROI – (C) Typical ROIs at different Gleason score s– (D) Comparison by pathologist and agreement percentages (intra-pathologist and Inferences versus Ground Truth).

| Image quality criteria | Comment |
|---|---|
| Nuclear components | The chromatin is distinct, with blue/purple staining |
| | The nucleolus is well-contrasted, dark blue-purple |
| | The nuclear membrane is clearly defined |
| Morphological characteristics of the cytoplasm | Good contrast compared to the extracellular matrix. |
| | Eosinophilic granules are well-defined, with red-orange staining |
| Extracellular matrix components | Erythrocytes are bright red |
| | Collagen is yellow-orange in staining (with H&E); the fibrillar structure is apparent |

Table 1. Image quality criteria adopted for preclinical evaluation done on prostate tissue sections.

| Rating | Comment |
|---|---|
| 4/4 (Optimal) | Excellent, near perfect |
| 3/4 (Good) | Good, with a few areas that could be improved |
| 2/4 (Bad) | Average quality, but sufficient to establish a diagnosis or prognosis. |
| 1/4 (Unsuitable) | Insufficient, which may result in diagnostic or prognostic errors. |

Table 2. Meaning of the selected rating

| Rating | Gleason Scores | ISUP | Meaning |
|---|---|---|---|
| 0 | Healthy | 0 | No cancer |
| 1 | IHC | NA | Doubt on the tissue integraty - IHC required to rise the doubt |
| 6 | Grade 6 (3+3) | I | Well-differentiated, uniform glands, minimal risk of aggressive behavior |
| 7 | Grade 7 (3+4) | II | Predominantly well-differentiated or moderately differentiated with focal moderate differentiation |
| 7 | Grade 7 (4+3) | III | Predominantly moderately or poorly differentiated with minor component of well differentiated glands |
| 8 | Grade 8 (4+4) | IV | Poorly differentiated, cribriform or fused glands, significant risk of aggressive behavior |
| 9 | Grade (4+5 or 5+4) | V | Very high risk: Poorly differentiated or undifferentiated, high likelihood of metastasis. |
| 10 | Grade 10 (5+5) | | |

Table 3. Rating used to perform the biostatistics

| Inputs | M | Model | Tissue | SOST | # test patches | SSIM (↑) | PSNR (↑) | LPIPS (↓) | BSA | Reference |
|---|---|---|---|---|---|---|---|---|---|---|
| IR + Brightfield | x20 | MUGAN | Prostate | yes | 13073 | 0.78 | 23.71 | 0.236 | yes | This paper |
| Autofluorescence | x40 | RegiStain | Autopsy lung | yes | ~96100 | 0.82 | 20.32 | - | - | (Li et al., 2024) |
| Autofluorescence | x40 | Brownian bridge diffusion | Lung | yes | ~2530 | 0.69 | 18.52 | 0.27 | - | (Zhang et al., 2025) |
| Brightfield | x40 | pix2pix | Murine Prostate | yes | ~90000 | 0.746 | 22.865 | - | yes | (Latonen et al., 2024) |
| Brightfield | x20 | pix2pix | Prostate | no | ~214960 | 0.902 | 22.82 | - | - | (Rana et al., 2020) |

Table 4. Comparison of the metric values obtained with our model with values reported for some works from the literature – M : magnification ; SOST : Strict out-of-sample testing ; BSA: Background subtraction applied

| Optimal image $Rating = 4\ (\geq 3)$ | $P_1$ | | $P_2$ | |
|---|---|---|---|---|
| | Ground truth | Digital staining | Ground truth | Digital staining |
| **Nucleus** | 87% (100%) | 67% (97%) | 90% (100%) | 81% (96%) |
| **Cytoplasm** | 81% (100%) | 72% (100%) | 90% (99%) | 72% (95%) |
| **Collagen** | 15% (99%) | 18% (99%) | 87% (100%) | 86% (100%) |

Table 5. Percentages of ROI considered as optimal (score = 4) by the two pathologists for the three cellular components : comparison between ground truth and digital staining – In parentheses, the percentage of ROIs with a score larger or equal to 3 (good)

|  | Nucleus | | Cytoplasm | | Collagen | |
| --- | --- | --- | --- | --- | --- | --- |
|  | $P_1$ | $P_2$ | $P_1$ | $P_2$ | $P_1$ | $P_2$ |
| $\chi^2\ p-value$ | 0.0012* | 0.0544 | 0.1121 | 0.0025* | 0.4693 | 0.8420 |

Table 6. Stain quality scoring – Statistical analysis with the $\chi^2$-test – * 5% significance level

| Two-sided test between Ground truth and Inferred | Gleason | |
| --- | --- | --- |
|  | $P_1$ | $P_2$ |
| $\chi^2\ p-value$ | 0.890 | 0.515 |

Table 7. Gleason grading – Statistical analysis with the $\chi^2$-test